\documentclass[aps,prd,amsmath,amssymb,twoside,showpacs,nofootinbib,preprintnumbers]{revtex4}

\usepackage[T1]{fontenc}

\newcommand{\defeq}{:=}
\newcommand{\im}{\mathrm{i}}  
\newcommand{\R}{\mathbb{R}}
\newcommand{\C}{\mathbb{C}}
\newcommand{\xd}{\mathrm{d}}
\newcommand{\xD}{\mathcal{D}} 
\newcommand{\cS}{\mathcal{S}}
\newcommand{\cH}{\mathcal{H}}
\newcommand{\cHc}{\mathcal{H}_\text{cyl}}
\newcommand{\tens}{\otimes}

\newcommand{\be}{\begin{equation}}
\newcommand{\ee}{\end{equation}}

\begin{document}

\title{Spatially asymptotic S-matrix from general boundary formulation}
\author{Daniele Colosi}\email{colosi@matmor.unam.mx}
\author{Robert Oeckl}\email{robert@matmor.unam.mx}
\affiliation{Instituto de Matem\'aticas, UNAM, Campus Morelia,
C.P.~58190, Morelia, Michoac\'an, Mexico}
\date{16 July 2008 (v2)}
\preprint{UNAM-IM-MOR-2008-1}

\begin{abstract}
We construct a new type of S-matrix in quantum field theory using the general boundary formulation. In contrast to the usual S-matrix the space of free asymptotic states is located at spatial rather than at temporal infinity. Hence, the new S-matrix applies to situations where interactions may remain important at all times, but become negligible with distance. We show that the new S-matrix is equivalent to the usual one in situations where both apply. This equivalence is mediated by an isomorphism between the respective asymptotic state spaces that we construct.
We introduce coherent states that allow us to obtain explicit expressions for the new S-matrix.
In our formalism crossing symmetry becomes a manifest rather than a derived feature of the S-matrix.
\end{abstract}

\maketitle

\section{Introduction}

The general boundary formulation (GBF) \cite{Oe:boundary,Oe:GBQFT,Oe:probgbf} is a formulation of quantum theory that takes explicit account of spacetime. While in the standard formulation of quantum theory time does play an essential role, space only enters as a secondary concept as an ingredient of specific theories. Ironically, it is precisely this ``more'' in structure of going from time to spacetime that allows one to use ``less'' of it. Namely, while in the standard formulation the metric nature of time is essential, in the GBF only the topological structure of spacetime is indispensable. The latter turns out to be sufficient to formulate such fundamental concepts as probability conservation \cite{Oe:GBQFT}. Hence, the GBF is naturally suitable to formulate quantum theories without metric background, such as quantum gravity.
Furthermore, it allows to avoid other problems that arise in many approaches to quantizing gravity that are manifestly free of a background metric, such as the problem of locality. In quantum field theory the fact that we may choose to describe a given system only, without worrying about the rest of the universe, rests on principles such as causality and cluster decomposition. These in turn depend on the background metric of spacetime. Hence, without such a metric there is a priori no way to separate a system from the rest of the universe. In the GBF such a separation is achieved from the outset simply because one deals with the physics of spacetime regions that are explicitly separated from the rest of the universe through a boundary.

In the GBF state spaces that describe the objects of the theory are associated with boundaries of regions of spacetime (or components of these boundaries). One might think of such a state space as encoding the information that can potentially be exchanged between the region and the rest of the universe through the boundary. Amplitudes are associated with such regions and boundary states. These encode physical processes within the region and allow to calculate probabilities associated with measurements performed on the region. The standard state spaces and transition amplitudes are recovered when the spacetime region in question is a time interval times all of space. The boundary state space then splits into a tensor product of the initial state space with the final one, associated to initial and final boundary components respectively. The amplitude corresponding to this region becomes just the usual transition amplitude.

To emphasize it again, the GBF is merely a theoretical framework and not in itself any specific quantum theory. Rather, it is a specific definition of what constitutes a quantum theory and how predictions are extracted in principle from its ingredients. What kind of spacetime structures (regions and hypersurfaces) enter into the GBF depends on the theory under consideration. The minimum required are topological manifolds of a given dimension, for an example of this type see \cite{Oe:2dqym}. Slightly more structure would be provided by differentiable manifolds. One might reasonably assume that this is the relevant geometric setting for a quantum theory of gravity. Lorentzian manifolds and, more specifically, submanifolds of Minkowski space are the relevant ingredient for ordinary quantum field theory. For conformal field theory one would have manifolds with conformal structure etc.

Providing a viable framework for quantum gravity is one of the main motivations for the GBF. The program associated with this goal consists of establishing and developing the GBF in the context of known and tested quantum theories as well as with simplified models incorporating key features expected of a quantum theory of gravity.
A key conjecture of this program is that ordinary quantum field theory can be extended to fit into this framework for a sufficiently interesting class of admissible spacetime regions and hypersurfaces. We refer to this in the following as the \emph{extensibility conjecture}. That quantum field theory fits into the GBF if we restrict to equal-time hyperplanes and time-interval regions is essentially trivial. The emphasis is hence on the word ``interesting''.
While it is technically difficult to deal with generic hypersurfaces and regions, even very special and highly symmetric geometries can serve to demonstrate the power of the GBF if they differ from the standard geometry in sufficiently radical ways.

For example, an important restriction of the standard framework is that state spaces are only ever associated to spacelike (Cauchy) hypersurfaces. The more technical reason for this is that canonical quantization prescriptions usually rely on specific correspondences in the classical theory between data on such hypersurfaces and solutions in all of spacetime. The more conceptual reason for this is that in the probability interpretation of transition amplitudes we are used to think in terms of a strict temporal ordering between the prepared and the observed states. This might lead one to suspect that serious problems appear when trying to formulate the quantum theory with more general hypersurfaces, namely hypersurfaces that contain timelike parts.
It was shown in \cite{Oe:timelike} that these objections are unfounded. More precisely, it was shown for the Klein-Gordon quantum field theory how all the ingredients of the theory including vacuum, full state space, amplitudes etc.\ can be consistently defined on \emph{timelike} hyperplanes and associated interval-like regions. In crucial difference to the spacelike case, a state can not be generally labeled as incoming or outgoing. Rather each particle in a general multi-particle state may individually be incoming or outgoing.

This example still shares an important feature with the standard framework. Although the temporal ordering between the participating states is lost, an amplitude can still be thought of as a \emph{transition} amplitude between one state and another. In geometric terms this is because the region with which the amplitude is associated has a boundary which decomposes into two connected components, namely the hypersurfaces which carry the states.
A further step was hence taken in \cite{Oe:KGtl} where it was shown explicitly that the division into two state spaces associated to different boundary components can be given up. The geometry used was the solid hypercylinder $\R\times B^3_R$, i.e., a three-ball of radius $R$ in space extended over all of time. The boundary $\R\times S^2_R$ of this region is \emph{connected}, implying that the associated state space does not decompose into a tensor product of components associated to subspaces of this boundary. Nevertheless, physical sense can be made of amplitudes associated with states on this hypercylinder and probabilities be extracted.

The extensions of quantum field theory described so far were only elaborated in the free theory. This is not surprising. Even in the standard framework we mostly do not know what the state space of an interacting quantum field theory is. However, we have a highly successful technique which allows us to say a lot about interacting quantum field theory without this knowledge: The S-matrix. Assuming that interactions are only relevant at intermediate times, one considers the transition amplitude between free states at an initial time and free states at a final time. One then takes the asymptotic limit of this amplitude where the initial time goes to $-\infty$ and the final time goes to $+\infty$. This yields the S-matrix.

A convincing argument for the validity of the extensibility conjecture in interacting quantum field theory can be made, if we manage to construct an asymptotic amplitude starting from an interesting non-conventional geometry and show its physical equivalence to the S-matrix. This is precisely what we announced in a previous letter \cite{CoOe:spsmatrix} and detail in this paper. The geometry we choose is that of the aforementioned hypercylinder $\R\times S^2$ as it is not only timelike, but also connected and exhibits most of the new features of the GBF in this context.
Physically, the starting assumption is that interactions can be neglected outside a certain finite region of space. However, no assumption needs to be made about interactions being negligible at any specific time. We consider then the amplitude associated with the solid hypercylinder $\R\times B^3_R$ of sufficiently large radius $R$ for a free state on its boundary $\R\times S^2_R$. The sought for asymptotic amplitude arises in the limit $R\to\infty$.
We go on to show that there is an isomorphism of Hilbert spaces identifying the usual (temporal) asymptotic state space of the S-matrix with the (spatial) asymptotic state space of the hypercylinder in the infinite radius limit. Under this mapping the S-matrix and the asymptotic hypercylinder amplitude become identical.

The paper is organized as follows. In Section \ref{sec:basics} we recall some basic aspects of the massive Klein-Gordon field in Minkowski spacetime. In particular, we use the Feynman path integral quantization procedure combined with the Schr\"odinger representation and coherent states. In Section \ref{sec:smatrix} the S-matrix elements in the basis of the coherent states are computed in three steps: we first consider the free theory, then we study the interaction of the scalar field with an external source, and finally functional methods allow us to work out the S-matrix for the general interacting theory. While the resulting S-matrix is well known, the purpose of our derivation is to serve as a blueprint for the novel case of the asymptotic amplitude on the hypercylinder that is treated later. Also, we have not found the derivation in this form elsewhere and it might thus have some interest of its own.
In Section \ref{sec:gbf} we treat the Klein-Gordon field defined on an hypercylinder within the GBF. We recall the main features of the classical and the quantum theories studied in \cite{Oe:KGtl} adapting them to the needs of the present setting. We then proceed to introduce coherent states on the hypercylinder. In Section \ref{sec:hsmatrix} the asymptotic amplitude for the hypercylinder is derived in analogy to the same three steps performed in Section \ref{sec:smatrix}. In Section \ref{sec:compstates} we compare the expressions of the two asymptotic amplitudes obtained, showing them to be identical under a suitable isomorphism of the asymptotic state spaces.
We end by presenting conclusions and an outlook in Section~\ref{sec:conclude}. Appendix~\ref{sec:transcoh} describes the conversion of coherent states from the Fock representation to the Schr\"odinger representation. Appendix~\ref{sec:appB} contains useful formulas for spherical harmonics and different types of Bessel functions.

\section{Free Klein-Gordon field in Minkowski spacetime}
\label{sec:basics}

We start with the real massive Klein-Gordon quantum field theory in
Minkowski spacetime whose action in a region $M$ is given by
\be
S_{M,0}(\phi)=\frac{1}{2}\int_M \xd^4 x\, \left((\partial_0
\phi)(\partial_0 \phi)- \sum_{i\ge 1} (\partial_i \phi)(\partial_i
\phi) -m^2\phi^2\right) .
 \label{eq:freeact}
\ee
(The index $0$ indicates that we consider the free theory.)
The equation of motion is the Klein-Gordon equation
\be
\left(\Box+m^2\right) \phi = 0,
\ee
where $\Box=\partial_0^2- \sum_{i\ge 1}\partial_i^2$.

\subsection{Schr\"odinger-Feynman approach}

We use the Schr\"odinger representation for quantum states. That is,
quantum states are wave functions on the space of instantaneous field
configurations. While this is the usual approach in non-relativistic
quantum theory it is less frequently used in quantum field theory, but
see \cite{Jac:schroedinger,Hat:qft}. At the same time we use the
Feynman path integral to represent transition amplitudes. In contrast
to other quantization prescriptions, such as canonical quantization,
this setting generalizes readily to the general boundary formulation,
which we will use starting from Section~\ref{sec:gbf}.

We recall below basic elements of the Schr\"odinger-Feynman setting
applied to the Klein-Gordon quantum field theory. The inner product of
states is,
\be
 \langle \psi_2 | \psi_1\rangle
 =\int \xD\varphi\, \psi_1(\varphi)\overline{\psi_2(\varphi)},
\label{eq:ip1}
\ee
where the integral is over all field configurations.

The transition amplitude from the state of the system at time $t_1$
described by the wave function $\psi_1(\varphi_1)$ to the state of the
system at time $t_2$ described by the wave function $\psi_2(\varphi_2)$
takes the form
\be
\langle \psi_2|U_{[t_1,t_2]}|\psi_1\rangle =
\int \xD\varphi_1\,\xD\varphi_2\, \psi_{1}(\varphi_1)
\overline{\psi_{2}(\varphi_2)}\, Z_{[t_1,t_2]}(\varphi_1,\varphi_2),
\label{eq:tampl1}
\ee
where $U_{[t_1,t_2]}$ represents the time-evolution operator, and the
field propagator is given by
\be
Z_{[t_1,t_2]}(\varphi_1,\varphi_2)=
\int_{\substack{\phi|_{t_1}=\varphi_1\\ \phi|_{t_2}=\varphi_2}}
\xD\phi\, e^{\im S_{[t_1,t_2]}(\phi)}.
\label{eq:proppint}
\ee
In the case of the free theory determined by the action
(\ref{eq:freeact}) we can evaluate the associated propagator
$Z_{[t_1,t_2],0}$ by
shifting the integration variable by a classical solution
$\phi_\text{cl}$ that interpolates between $\varphi_1$ at $t_1$ and
$\varphi_2$ at $t_2$. Although the result can be readily found in
\cite{Hat:qft} or \cite{Oe:KGtl}, we repeat the derivation here since
it provides an instructive example for calculations in later sections
that follow the same pattern.
Explicitly,
\be
Z_{[t_1,t_2],0}(\varphi_1,\varphi_2) =
\int_{\substack{\phi|_{t_1}=\varphi_1\\ \phi|_{t_2}=\varphi_2}}
\xD\phi\, e^{\im S_{[t_1,t_2],0}(\phi)}
=  \int_{\substack{\phi|_{t_1}=0\\ \phi|_{t_2}=0}} \xD\phi\, e^{\im
  S_{[t_1,t_2],0}(\phi_{cl}+\phi)}
= N_{[t_1,t_2],0} \,  e^{\im S_{[t_1,t_2],0}(\phi_\text{cl})} ,
\label{eq:propfree}
\ee
where the normalization factor is formally given by
\be
N_{[t_1,t_2],0}=\int_{\substack{\phi|_{t_1}=0\\
\phi|_{t_2}=0}}
\xD\phi\, e^{\im S_{[t_1,t_2],0}(\phi)} .
\label{eq:n0}
\ee
$\phi_\text{cl}$ can be decomposed into positive and negative energy modes as
\be
\phi_{cl}(x, {t}) = e^{-i \omega {t}} \varphi^+(x) +  e^{i \omega {t}}
\varphi^-(x).
\label{eq:phi0}
\ee
Here the time variable ${t}$ belongs to the interval $[t_1,t_2]$ and
$\omega$ is the operator
\be
\omega\defeq\sqrt{-\sum_{i\ge 1}\partial_i^2+m^2}.
\ee
The configurations $\varphi^{\pm}$ are related to the configurations
on the boundary, $\varphi_1$ for ${t}=t_1$ and $\varphi_2$ for
${t}=t_2$, by the relation
\be
\left(\begin{matrix} \varphi_1 \\ \varphi_2 \end{matrix} \right) =
\left( \begin{matrix}
e^{-i \omega t_1} & e^{i \omega t_1} \\
e^{-i \omega t_2} & e^{i \omega t_2}
\end{matrix} \right)
\left(\begin{matrix} \varphi^+ \\ \varphi^- \end{matrix} \right).
\ee
We invert the matrix to express the configurations
$\varphi^{\pm}$ as function of the configurations $\varphi_1$ and
$\varphi_2$:
\be
\left(\begin{matrix} \varphi^+ \\ \varphi^- \end{matrix} \right)
=  \frac{1}{2i \sin \omega (t_2-t_1)}
\left( \begin{matrix}
e^{i \omega t_2} & -e^{i \omega t_1} \\
-e^{-i \omega t_2} & e^{-i \omega t_1}
\end{matrix} \right)
\left(\begin{matrix} \varphi_1 \\ \varphi_2 \end{matrix} \right).
\ee
Substituting this into (\ref{eq:phi0}), we get the expression for the
classical solution $\phi_\text{cl}$ in terms of $\varphi_1$ and
$\varphi_2$,
\be
\phi_\text{cl}(t,x)=
\frac{\sin\omega(t_2-t)}{\sin\omega(t_2-t_1)}\varphi_1(x)
+\frac{\sin\omega(t-t_1)}{\sin\omega(t_2-t_1)}\varphi_2(x) .
\label{eq:clsol-boundary}
\ee
This allows in turn to evaluate the field propagator,
\be
 Z_{[t_1,t_2],0}(\varphi_1,\varphi_2)=N_{[t_1,t_2],0}
 \exp\left(-\frac{1}{2}\int\xd^3 x\,
 \begin{pmatrix}\varphi_1 & \varphi_2\end{pmatrix} W_{[t_1,t_2]}
 \begin{pmatrix}\varphi_1\\ \varphi_2\end{pmatrix}\right),
 \label{eq:freeprop}
\ee
where $W_{[t_1,t_2]}$ is the operator valued $2\times 2$ matrix
\be
W_{[t_1,t_2]}=\frac{-\im\omega}{\sin\omega(t_2-t_1)}
\begin{pmatrix} \cos\omega(t_2-t_1) & -1\\
 -1 & \cos\omega(t_2-t_1)\end{pmatrix} .
\ee

The vacuum wave function is,
\be
 \psi_0(\varphi)=C
 \exp\left(-\frac{1}{2}\int\xd^3 x\,\varphi(x)(\omega\varphi)(x)\right) ,
\label{eq:vacuum}
\ee
where $C$ is a normalization factor.

\subsection{Coherent states}
\label{sec:coherent-state}

Pioneered by Glauber \cite{Gla:quantcoh} to study electromagnetic
correlation functions in quantum optics, coherent state techniques
are by now widely used in quantum field theory and particle
physics. The interest in coherent states comes mainly from the possibility
to construct quantum states that
correspond, in the limit where the number of field quanta is large, to
classical field configurations. We define
coherent states for the Klein-Gordon theory in Minkowski spacetime
following the conventions in \cite{ItZu:qft}. We
first adopt a Fock representation to express the coherent
states and then pass to the Schr\"odinger representation.

Indicating the vacuum state of the scalar field in the Fock
representation with $|0 \rangle$, a normalized coherent state takes
the form
\be
| \psi_\eta \rangle = C_\eta \exp \left( \int \frac{\xd ^3k}{(2
  \pi)^3 2 E} \, \eta(k) a^{\dagger}(k)\right) |0 \rangle,
\label{eq:coherentstate}
\ee
where $a^{\dagger}(k)$ is the creation operator for the field mode of
momentum $k$. $\eta$ is a complex function on momentum space that
characterizes the state. The normalization constant $C_\eta$ is given
by
\be
C_\eta=\exp \left(-\frac{1}{2}\int \frac{\xd ^3k}{(2 \pi)^3 2 E}\,
|\eta(k)|^2 \right)
\ee
so that the inner product of coherent states is given by
\be
\langle \psi_{\eta_2}|\psi_{\eta_1}\rangle
=\exp \bigg( \int \frac{\xd^3k}{(2 \pi)^3 2 E} \,
\left(  \eta_1(k)\, \overline{\eta_2(k)}  - \frac{1}{2} |\eta_1(k)|^2
 - \frac{1}{2} |\eta_2(k)|^2\right) \bigg) .
\label{eq:sproduct-cs}
\ee
We can write the resolution of the identity operator $I$ in terms of
coherent states as
\be
D^{-1} \int \xd \eta \, \xd \overline{\eta} \,
 |\psi_{\eta} \rangle \langle \psi_{\eta}| = I,
 \label{eq:cohcompl}
\ee
with
\be
 D=\int\xd\eta\,\xd\overline{\eta}\,
 \exp\left(-\int\frac{\xd^3 k}{(2\pi)^3 2 E}\, |\eta(k)|^2\right) .
\ee
The time evolution of a coherent state is given by
\be
e^{-\im H \Delta t}|\psi_\eta \rangle =|\psi_{\eta'}\rangle\quad
\text{with}\quad
\eta'(k)=e^{-\im E \Delta t} \eta(k).
\label{eq:cohtev}
\ee

The expansion of a coherent state in terms of multi-particle states
can be read off directly from (\ref{eq:coherentstate}),
\be
|\psi_\eta\rangle = C_\eta \sum_{n=0}^\infty \frac{1}{n!}
\int \frac{\xd ^3k_1}{(2
  \pi)^3 2 E_1} \,\cdots\,\int \frac{\xd ^3k_n}{(2
  \pi)^3 2 E_n}\,\eta(k_1)\cdots\eta(k_n)\,
|\psi_{k_1,\dots,k_n}\rangle
\label{eq:cohexpand}
\ee
In particular, the inner product between a coherent state determined
by $\eta$ and a state with particles of momenta $k_1,\dots,k_n$ is,
\be
 \langle \psi_{k_1,\dots,k_n} | \psi_\eta \rangle
 = C_\eta \, \eta(k_1)\cdots\eta(k_n) .
\label{eq:cohnip}
\ee

As shown in Appendix~\ref{sec:transcoh} the expression for the
coherent state in the Schr\"odinger representation is
\be
\psi_{\eta} (\varphi) = K_{\eta} \exp \left( \int
\frac{\xd^3 x\,\xd^3 k}{(2 \pi)^3} \, \eta(k) \, e^{\im k x} \,
\varphi(x)\right)\, \psi_0(\varphi) ,
\ee
with $\psi_0(\varphi)$ being the wave function of the vacuum
state. The normalization factor $K_{\eta}$, different from the
normalization factor $C_\eta$ in the Fock representation is given by
\be
K_{\eta}= \exp \left( - \frac{1}{2}\int  \frac{\xd^3 k}{(2 \pi)^3
2 E}  \left(  \eta(k)\eta(-k)
+ | {{\eta}(k)}|^2 \right)\right).
\label{eq:cohnorm}
\ee

\section{Conventional S-matrix}
\label{sec:smatrix}

The S-matrix is the standard tool to calculate probabilities of
scattering processes in quantum field theory. One assumes that
interactions can be neglected at very early and at very late times,
where states are typically considered as consisting of a collection of
free particles. Thus, to model these particles one uses the state
space of the theory without interaction. Transition amplitudes between
such free states at initial time $t_1$ and at final time $t_2$
can be calculated with the interaction switched on at intermediate
times. The S-matrix is then the asymptotic limit of these transition
amplitudes for $t_1\to -\infty$ and $t_2\to +\infty$.

In this section we compute the elements of the S-matrix in
three different cases. First we consider the free Klein-Gordon theory,
then we study the interacting theory where the interaction is given by
a source term, and we finally treat general interactions by means
of functional derivatives of the result obtained with the source
interaction.

Although the result is standard, our derivation differs from
standard textbook treatments. What is more, it provides a blueprint
to the subsequent derivation of a new type of S-matrix in
Section~\ref{sec:hsmatrix}. Our treatment
is similar in spirit to the one using the holomorphic
representation \cite{FaSl:gaugeqft}. However, instead of dealing with
propagation kernels we directly deal with transition amplitudes, using
the coherent states of Section~\ref{sec:coherent-state}.

\subsection{Free theory}

In order to make sense of the limiting procedure involved in defining
the S-matrix we switch to the interaction (or Dirac) picture. That is,
we identify states at different times if they  are related by time
evolution in the free theory. To make this more transparent we add
a label $t$ to the state when appropriate, specifying at which
time it is evaluated. Recalling the free time evolution
(\ref{eq:cohtev}) we obtain the wave function
\be
\psi_{t,\eta} (\varphi) = K_{t,\eta} \exp \left( \int
\frac{\xd^3 x\,\xd^3 k}{(2 \pi)^3} \, \eta(k) \, e^{-\im(E t-k x)} \,
\varphi(x)\right)\, \psi_0(\varphi) ,
\label{coherent-states-Schrodinger}
\ee
for the coherent state at time $t$ which at time $0$ coincides with
(\ref{eq:coherentstate}). Note that the normalization factor
(\ref{eq:cohnorm}) is now also time dependent and takes the form
\be
K_{t,\eta}= \exp \left( - \frac{1}{2}\int  \frac{\xd^3 k}{(2 \pi)^3
2 E}  \left(  e^{- 2\im E t} {\eta}(k)\eta(-k)
+ | {{\eta}(k)}|^2 \right)\right).
\label{eq:ipnorm}
\ee

The transition amplitude in the free theory in terms of interaction
picture states is by construction independent of the initial and final
time and equal to the inner product (\ref{eq:sproduct-cs}). In
particular, we can let the initial time go to $-\infty$ and the final
time go to $+\infty$ and think of these transition amplitudes as the
elements of the S-matrix $\cS_0$ of the free theory,
\be
\langle \psi_{\eta_2}|\cS_0|\psi_{\eta_1}\rangle
=\lim_{\substack{t_1\to-\infty\\ t_2\to+\infty}}
\langle \psi_{t_2,\eta_2}|U_{[t_2,t_1],0}|\psi_{t_1,\eta_1}\rangle
=\langle \psi_{\eta_2}|\psi_{\eta_1}\rangle .
\label{eq:smfree}
\ee
Here, we have denoted with $U_{[t_2,t_1],0}$ the evolution operator of
the free theory between the times $t_1$ and $t_2$.

\subsection{Theory with source}
\label{sec:theorysrc}

We consider in this section the Klein-Gordon theory interacting with a
source field $\mu$. In a spacetime region $M$ the new action takes the
form
\be
S_{M,\mu}(\phi)=S_{M,0}(\phi)+\int_M \xd^4 x\,  \mu(x) \phi(x),
\label{eq:actionsrc}
\ee
where $S_{M,0}$ is the free action (\ref{eq:freeact}). Here, $M$ will
be determined by a time interval $[t_1,t_2]$ and we assume the source
field to vanish outside this time interval.

The path integral (\ref{eq:proppint}) determining the propagator
$Z_{[t_1,t_2],\mu}$ for the theory with source can be evaluated in the
same way (\ref{eq:propfree}) as in the free theory. That is, we shift
the integration variable by a classical solution $\phi_\text{cl}$ of
the free theory (not the one with source) interpolating between
$\varphi_1$ at $t_1$ and $\varphi_2$ at $t_2$ to obtain
\be
Z_{[t_1,t_2],\mu}(\varphi_1,\varphi_2)=N_{[t_1,t_2],\mu}\,
e^{\im S_{[t_1,t_2],\mu}(\phi_\text{cl})} .
\label{eq:propsrc1}
\ee
The normalization factor is formally,
\be
N_{[t_1,t_2],\mu}=\int_{\substack{\phi|_{t_1}=0\\
\phi|_{t_2}=0}}
\xD\phi\, e^{\im S_{[t_1,t_2],\mu}(\phi)} .
\label{eq:nmu}
\ee
Separating the free propagator in (\ref{eq:propsrc1}) yields,
\be
 Z_{[t_1,t_2],\mu}(\varphi_1,\varphi_2)=
 \frac{N_{[t_1,t_2],\mu}}{N_{[t_1,t_2],0}}
  Z_{[t_1,t_2],0}(\varphi_1,\varphi_2)
 \exp\left(\im \int\xd^4 x\,  \mu(x) \phi_\text{cl}(x)\right) .
 \label{eq:propsrc2}
\ee
Using the decomposition (\ref{eq:clsol-boundary}) we can rewrite this
as,
\be
 Z_{[t_1,t_2],\mu}(\varphi_1,\varphi_2)=
 \frac{N_{[t_1,t_2],\mu}}{N_{[t_1,t_2],0}}
  Z_{[t_1,t_2],0}(\varphi_1,\varphi_2)
 \exp\left(\int\xd^3 x\,\left(\mu_1(x) \varphi_1(x)
 +\mu_2(x) \varphi_2(x)\right)\right) .
 \label{eq:propsource}
\ee
Here, $\mu_1$ and $\mu_2$ are defined as follows,
\be
 \mu_1(x)  \defeq  \im \int_{t_1}^{t_2}\xd t\,
 \frac{\sin\omega(t_2-t)}{\sin\omega(t_2-t_1)}\mu(t,x), \qquad
\mu_2(x)  \defeq \im \int_{t_1}^{t_2}\xd t\,
 \frac{\sin\omega(t-t_1)}{\sin\omega(t_2-t_1)}\mu(t,x) \label{eq:mu}.
\ee
Denoting the associated time-evolution operator
by $U_{[t_1,t_2],\mu}$, the transition amplitude between coherent
states is,
\begin{multline}
 \langle \psi_{t_2,\eta_2}|U_{[t_1,t_2],\mu}| \psi_{t_1,\eta_1}\rangle
= K_{t_1,\eta_1}\overline{K_{t_2,\eta_2}} \int \xD\varphi_1\,\xD\varphi_2\\
\exp \left( \int \frac{\xd^3 x\,\xd^3 k}{(2 \pi)^3} \,
\left(\eta_1(k) \,
e^{-\im(E t_1-k x)} \, \varphi_1(x)+\overline{\eta_2(k)} \,
e^{\im(E t_2-k x)} \, \varphi_2(x)\right)\right)\,
\psi_0(\varphi_1)\,\overline{\psi_0(\varphi_2)}\,
Z_{[t_1,t_2],\mu}(\varphi_1,\varphi_2) .
\label{eq:tampl-src2}
\end{multline}
To evaluate this expression, we observe that we can combine the
exponential factor appearing in (\ref{eq:propsource}) with the
exponential in the above formula. We can then reduce the transition
amplitude (up to normalization) to one of the free theory, but between
modified coherent states defined by complex functions $\tilde{\eta}_1$ and $\tilde{\eta}_2$ given by
\be
\tilde{\eta}_1(k)\defeq \eta_1(k)
+ \int \xd^3 x\, e^{\im(E t_1-k x)} \mu_1(x) \quad \text{and} \quad
\tilde{\eta}_2(k)\defeq\eta_2(k)
+ \int \xd^3 x\, e^{\im(E t_2-k x)} \overline{\mu_2(x)}.
\ee
We obtain
\be
 \langle \psi_{t_2,\eta_2}|U_{[t_1,t_2],\mu}|
 \psi_{t_1,\eta_1}\rangle  = \langle\psi_{\tilde{\eta}_2}|
\psi_{\tilde{\eta}_1}\rangle  \,
\frac{N_{[t_1,t_2],\mu}\, K_{t_1,\eta_1}\,
\overline{K_{t_2,\eta_2}}}{N_{[t_1,t_2],0}\, K_{t_1,\hat{\eta}_1}\,
   \overline{K_{t_2,\hat{\eta}_2}}}.
\label{eq:tampl-src3}
\ee
Using (\ref{eq:sproduct-cs}) to express the inner product between the modified coherent states and (\ref{eq:ipnorm}) for the state normalization factors, equation (\ref{eq:tampl-src3}) yields
\be
\langle \psi_{t_2,\eta_2}|U_{[t_1,t_2],\mu}| \psi_{t_1,\eta_1}\rangle
= \langle\psi_{\eta_2}| \psi_{\eta_1}\rangle \,
\frac{N_{[t_1,t_2],\mu}}{N_{[t_1,t_2],0}} \exp\left(\im\int\xd^4 x\, \mu(x) \hat{\eta}(x)\right) \exp\left(\frac{\im}{2}\int \xd^4 x\,\mu(x)\beta(x)\right),
\label{eq:tampl-src4}
\ee
where $\hat{\eta}$ is the complex classical solution of the Klein-Gordon equation
determined by $\eta_1$ and $\eta_2$ via
\be
\hat{\eta}(t,x)=
\int \frac{\xd^3 k}{(2\pi)^3 2 E}
\left(\eta_1(k) e^{-\im (E t- k x)}
+\overline{\eta_2(k)} e^{\im(E t- k x)}\right) .
\label{eq:classcoh}
\ee
On the other hand, any bounded solution of the Klein-Gordon equation can be expanded in this way. Hence, this establishes a one-to-one correspondence between pairs of coherent states parametrized by pairs of functions $(\eta_1,\eta_2)$ and solutions $\hat{\eta}$.

The function $\beta(x)$ appearing in the last exponential of (\ref{eq:tampl-src4}) is a solution of the Klein-Gordon equation whose Fourier
decomposition is given by
\be
\beta(t,k) = \int_{t_1}^{t_2} \frac{\xd \tau}{2 E}  \left( \im\, e^{ -\im E
    (\tau -t)}+ \, \frac{2\,\sin (E (t-t_1)) \, \sin
  (E(t_2-\tau))}{\sin (E (t_2-t_1))} \right) \, \mu(\tau,k),
\label{eq:beta}
\ee
where
\be
 \beta(t,k)= 2 E\int \xd^3 x\,\beta(t,x) e^{-\im k x},\quad\text{and}\quad
 \mu(t,k)= 2 E\int \xd^3 x\,\mu(t,x) e^{-\im k x} .
\ee

It remains to evaluate the propagator normalization factor
(\ref{eq:nmu}) and combine the result with the other terms in (\ref{eq:tampl-src3}). It is
possible to relate this normalization factor $N_{[t_1,t_2],\mu}$ to
that of the
free theory $N_{[t_1,t_2],0}$ defined in (\ref{eq:n0}) by shifting the
integration variable in (\ref{eq:nmu}) by a particular function
$\alpha$,
\be
\int_{\substack{\phi|_{t_1}=0\\ \phi|_{t_2}=0}} \xD\phi \;
e^{\im S_{[t_1,t_2],\mu}(\phi)} = \exp\left(\frac{\im}{2}\int \xd^4
x\,\mu(x)\alpha(x)\right)
\int_{\substack{\phi|_{t_1}=0\\ \phi|_{t_2}=0}} \xD\phi \; e^{\im
  S_{[t_1,t_2],0}(\phi)}.
\label{eq:propnormsrc}
\ee
That is,
\be
 \frac{N_{[t_1,t_2],\mu}}{N_{[t_1,t_2],0}}=\exp\left(\frac{\im}{2}\int \xd^4
x\,\mu(x)\alpha(x)\right),
\label{eq:alphafac}
\ee
where the function $\alpha$ is a solution of the inhomogeneous Klein-Gordon
equation
\be
(\Box+ m^2)\alpha(t,x)=\mu(t,x),
\label{eq:iKG}
\ee
in the spacetime region $[t_1,t_2]\times\R^3$ and with the boundary
conditions $\alpha(t_1,x)=0$ and $\alpha(t_2,x)=0$ for all
$x\in\R^3$. It will be convenient to work in momentum space and
consider the Fourier components of $\alpha$,
\be
\alpha(t,k) = 2 E \int \xd^3 x \, \alpha(t,x)\, e^{- \im kx}.
\ee
Equation (\ref{eq:iKG}) implies
\be
(\partial_0^2 + E^2) \alpha(t,k) = \mu(t,k),
\ee
with $E^2= k^2 + m^2$. The solution is easily found and can be written
in the form
\be
\alpha(t,k) = - \int_{t_1}^{t_2} \frac{\xd \tau}{2 E}  \left( \theta(t-\tau) \,
2 \sin (E (\tau -t)) +  \frac{2\,\sin (E (t-t_1)) \, \sin (E
  (t_2-\tau))}{\sin (E (t_2-t_1))} \right) \, \mu(\tau,k).
\label{eq:alpha}
\ee
where $\theta(t)$ is the step function
\be
\theta(t)= \left\{ \begin{matrix} 1 & \ \ \hbox{if} \ \ t>0\\
0 & \ \ \hbox{if} \ \ t<0.
\end{matrix}\right.
\label{eq:theta}
\ee

Combining the exponential factor containing $\beta$ in (\ref{eq:tampl-src4}) with
(\ref{eq:alphafac})
amounts to summing the solution $\beta$ given by (\ref{eq:beta}) and
$\alpha$ given by (\ref{eq:alpha}),
\be
\gamma(t,k) \defeq \alpha(t,k) +  \beta(t,k) =  \int_{t_1}^{t_2}
\frac{\xd \tau}{2 E}, \im\left( \theta(t -\tau) e^{\im E (\tau-t)}
  + \theta(\tau -t) e^{-\im E (\tau-t)}  \right) \,
\mu(\tau,k),
\label{eq:gamma}
\ee
Since $\alpha$ solves the
inhomogeneous equation (\ref{eq:iKG}) and $\beta$ the homogeneous
equation the sum $\gamma$ solves the inhomogeneous equation
(\ref{eq:iKG}) as well, but with different boundary conditions than
$\alpha$. Indeed, we can read off the boundary conditions from
(\ref{eq:gamma}):
\be\begin{split}
\text{for} \  \ t<t_1, \  \ \gamma(t,k) = \frac{e^{\im E t}}{2 E}
 \int_{t_1}^{t_2} \xd \tau
\, \im\, e^{ - \im E \tau} \, \mu(\tau,k),\\ 
\text{for} \  \ t>t_2, \  \  \gamma(t,k) = \frac{e^{-\im E t}}{2 E}
 \int_{t_1}^{t_2} \xd \tau \, \im\, e^{  \im E \tau}  \,
\mu(\tau,k).
\label{eq:gammabdy}
\end{split}\ee
Namely, the function $\gamma$ contains only
negative energy modes at early times ($t<t_1$) and positive energy
modes at late times ($t>t_2$). We recognize these as the \emph{Feynman
boundary conditions}. Thus, we can write
$\gamma$ in the following integral form,
\be
\gamma(x)  =  \int  \xd^4 x'  \, G_F(x,x')\mu(x'),
\label{eq:gammafeyn}
\ee 
where $G_F$ is the Feynman propagator normalized such that
$(\Box_x+m^2)G_F(x,x')=\delta^4(x-x')$.

Now inserting
(\ref{eq:alphafac}) into (\ref{eq:tampl-src4}) and using
(\ref{eq:gamma}) as well as (\ref{eq:gammafeyn}) we obtain for the
transition amplitude,
\be
\langle \psi_{t_2,\eta_2}|U_{[t_1,t_2],\mu}|
\psi_{t_1,\eta_1}\rangle=\langle\psi_{\eta_2}|\psi_{\eta_1}\rangle
\exp\left(\im\int\xd^4 x\, \mu(x) \hat{\eta}(x)\right)
\exp\left(\frac{\im}{2}\int \xd^4 x\, \xd^4 x' \mu(x)G_F(x,x')\mu(x')\right).
\label{eq:tampl-srcint}
\ee
This expression is independent of the times $t_1$ and $t_2$ as long as
the support of $\mu$ vanishes outside the interval $[t_1,t_2]$. Thus, we
can take the limits $t_{1} \rightarrow - \infty$ and $t_{2}
\rightarrow + \infty$, subsequently lift the restriction on the
support of $\mu$, and interpret the result as the S-matrix for the
theory with the source interaction,
\be
\langle \psi_{\eta_2}|\cS_{\mu}|
\psi_{\eta_1}\rangle=\langle\psi_{\eta_2}|\cS_0|\psi_{\eta_1}\rangle
\exp\left(\im\int\xd^4 x\, \mu(x) \hat{\eta}(x)\right)
\exp\left(\frac{\im}{2}\int \xd^4 x\, \xd^4 x' \mu(x)G_F(x,x')\mu(x')\right).
\label{eq:smsrc}
\ee

\subsection{General interactions}
\label{sec:generalint}

The result of the previous section can be combined with functional
derivative techniques to work out the S-matrix in the case of a
general interaction. The action of the scalar field with an arbitrary
potential $V$ in the spacetime region $M$ can be written as
\be
S_{M,V}(\phi)=S_{M,0}(\phi)+\int_M \xd^4 x\, V(x,\phi(x)).
\label{eq:actionpot}
\ee
We notice the usual functional identity,
\be
\exp\left(\im S_{M,V}(\phi)\right)= \exp\left(\im\int_M \xd^4 x\,
V\left(x,-\im \frac{\partial}{\partial
\mu(x)}\right)\right) \exp\left(\im
S_{M,\mu}(\phi)\right)\bigg|_{\mu=0},
\label{eq:actionfunc}
\ee
where $S_{M,\mu}$ is the action in the presence of a source interaction,
defined in (\ref{eq:actionsrc}).

As above, at first the spacetime region of
interest is determined by a time interval $[t_1,t_2]$
and we assume that the
interaction vanishes outside of it,
\be
V((t,x),\phi(t,x)) = 0, \forall x\in\R^3, \forall t \notin [t_1,t_2].
\label{eq:vvanish}
\ee
Inserting (\ref{eq:actionfunc}) into the path integral of the
propagator (\ref{eq:proppint}), we observe that we can move the exponential
containing the functional derivative out of the integral to the
front. In fact, we can
repeat all steps identically as for the theory with source, but with
the functional derivative term in front. Hence, the transition
amplitude can be read off from (\ref{eq:tampl-srcint}) in combination
with (\ref{eq:actionfunc}),
\be
 \langle \psi_{t_2,2}|U_{[t_1,t_2],V}| \psi_{t_1,1}\rangle
 = \exp\left(\im\int\xd^4 x\,
  V\left(x,-\im\frac{\partial}{\partial \mu(x)}\right)\right)\\
 \langle \psi_{t_2,2}|U_{[t_1,t_2],\mu}| \psi_{t_1,1}\rangle
 \bigg|_{\mu=0} .
\label{eq:aint}
\ee
In this expression a restriction to coherent states is not necessary,
so we have written it for general states.
We now recall that the transition amplitude (\ref{eq:tampl-srcint})
does not actually depend on the times $t_1$ and $t_2$, but is
identical to the S-matrix of the theory with source. Hence, expression
(\ref{eq:aint}) also does not depend on $t_1$ and $t_2$ and the
limits $t_1\to-\infty$ and $t_2\to\infty$ are trivial. We can lift the
restriction (\ref{eq:vvanish}) on $V$ and obtain the S-matrix,
\be
  \langle \psi_{2}|\cS_V| \psi_{1}\rangle
 = \exp\left(\im\int\xd^4 x\,
  V\left(x,-\im\frac{\partial}{\partial \mu(x)}\right)\right)\\
 \langle \psi_{2}|\cS_\mu| \psi_{1}\rangle
 \bigg|_{\mu=0} .
\label{eq:smint}
\ee

\section{General boundaries and the hypercylinder}
\label{sec:gbf}

The general boundary formulation of quantum theory posits that we can
assign physically meaningful
amplitudes not only to transitions between an initial and a final
spacelike hypersurface, but to more general bounded regions of
spacetime. Similarly, we can associate states (and state spaces) not
only to spacelike hypersurfaces, but to more general
hypersurfaces. A key conjecture in this context is that standard
quantum field theory admits such a generalization -- for a certain
class of regions and hypersurfaces that are yet to be determined. In
fact, the present paper may be viewed as proving that conjecture for a
certain very limited class of regions and hypersurfaces. Nevertheless,
this class is very different from the usual time-interval regions and
equal-time hypersurfaces and hence points to more general geometries.

We follow the Schr\"odinger-Feynman approach to quantization above,
adapting it to the more general context. Hence,
the state space $\cH_\Sigma$ for a hypersurface $\Sigma$ should be
the space of functions on field configurations $K_\Sigma$ on
$\Sigma$. We write the inner product there (naively) as
\be
 \langle \psi_2 | \psi_1\rangle
  = \int_{K_\Sigma}\xD\varphi\,
  \psi_1(\varphi)\overline{\psi_2(\varphi)} .
 \label{eq:iprod}
\ee
The amplitude for a region $M$ and a state $\psi$ in the
state space $\cH_{\partial M}$ associated to the boundary $\partial M$
of $M$ is the integral
\be
\rho_M(\psi) = \int_{K_{\Sigma}} \xD\varphi \, \psi(\varphi)
Z_M(\varphi),
\label{amplitude}
\ee
where the hypersurface $\Sigma=\partial M$ represents the boundary of $M$.
The quantity $Z_M$ is the propagator given by the Feynman path integral,
\be
Z_{M}(\varphi)= \int_{K_M, \phi|_{\Sigma}=\varphi} \xD\phi\,
 e^{\im S_{M}(\phi)}, \  \ \forall \varphi \in K_{\Sigma} .
\label{prop}
\ee
Here the integral is over the space $K_M$ of field configurations
$\phi$ in the interior of $M$ such that $\phi$ agrees with $\varphi$
on the boundary $\Sigma$.

When $\Sigma$ is an equal-time hypersurface, the inner product
(\ref{eq:iprod}) becomes (\ref{eq:ip1}). Similarly,
when $M=[t_1,t_2]\times \R^3$, i.e., it is a time-interval times all
of space in Minkowski spacetime, its boundary $\partial M$
consists of the disjoint union of two equal-time hypersurfaces, $\Sigma_1$
defined by the time $t_1$ and $\Sigma_2$ defined by $t_2$. Then, the
amplitude (\ref{amplitude}) specializes to the transition amplitude
(\ref{eq:tampl1}) and the propagator (\ref{eq:proppint}) to (\ref{prop}).
In what follows, we shall be interested in a very
different type of geometry.

\subsection{Klein-Gordon theory on the hypercylinder}
\label{sec:hypercyl}

\subsubsection{Classical theory}
\label{sec:hypclassical}

Consider a sphere of radius $R$ centered at the origin in space and
extended to all of time. We refer to this hypersurface $\R\times
S^2_R$ in Minkowski space as the \emph{hypercylinder} of radius
$R$, and to its interior $\R\times B^3_R$ as the \emph{solid
hypercylinder} of radius $R$. To describe fields on its boundary or in
its interior it is convenient to use spherical coordinates in
space, defined by three parameters: $\theta \in [0, \pi[$, $\phi \in
[0,2 \pi[$ and $r \in [0, \infty[$. Bounded solutions of the
Klein-Gordon equation may be expanded in products of
spherical harmonics, spherical Bessel functions and
exponentials. Consider a general expansion of the form
\be
\phi(t,r, \Omega) = \int_{-\infty}^{\infty} \xd E
 \sum_{l=0}^{\infty}\sum_{m=-l}^{l}
\alpha_{l,m}(E) e^{-iEt} f_l(E,r)Y_l^m (\Omega),
\label{eq:clkg}
\ee
where $f_l$ denotes a certain kind of spherical Bessel function to be
discussed below. $\Omega$ is a collective
notation for the angle coordinates $(\theta, \phi)$. 
$Y_l^m$ denotes the spherical harmonic. See Appendix~\ref{sec:appB}
for their definition and properties.
The spherical harmonics satisfy the differential equation
\be
(\Delta_{\Omega}Y_l^m)(\Omega) = - \frac{l(l+1)}{r^2} Y_l^m(\Omega), \
\ \hbox{with} \ \ \Delta_{\Omega}= \frac{\cos \theta}{r^2 \sin \theta}
\partial_{\theta}+ \frac{1}{r^2} \partial_{\theta}^2 +
\frac{1}{r^2 \sin^2 \theta} \partial_{\phi}^2.
\label{eq:diffY}
\ee

For (\ref{eq:clkg}) to solve the Klein-Gordon equation the function
$f_l$ must satisfy the differential equation,
\be
(\Delta_r f_l) (E,r) = \left(-E^2 + m^2 + \frac{l(l+1)}{r^2} \right)
 f_l(E,r),
\  \ \hbox{with} \ \ \Delta_r= \frac{2}{r} \partial_r + \partial_r^2.
\label{eq:diffbessel}
\ee
If $E^2\ge m^2$, this is solved by the \emph{spherical Bessel
functions} of the first kind $j_l$ and of the second kind $n_l$,
setting $f_l(E,r)=j_l(r\sqrt{E^2-m^2})$ or
$f_l(E,r)=n_l(r\sqrt{E^2-m^2})$ respectively. Both, $j_l$ and $n_l$
are real. If $E^2\le m^2$, this
is solved by the \emph{modified spherical Bessel functions} of the
first kind $i_l^+$ and of the second kind $i_l^-$, setting
$f_l(E,r)=i_l^+(r\sqrt{m^2-E^2})$ or $f_l(E,r)=i_l^-(r\sqrt{m^2-E^2})$
respectively. For the definition of the different types of spherical
Bessel functions and their relations, see Appendix~\ref{sec:appB}. 

Not all types of Bessel functions lead to solutions of
the Klein-Gordon equation that are well-defined and bounded
everywhere:
\begin{itemize}
\item The ordinary and modified spherical Bessel functions of the second kind, $n_l$ and $i_l^-$ respectively are singular at the origin. So the associated solutions are singular on the time axis.
\item The spherical Bessel functions of the first and second kind, $j_l$ and $n_l$ respectively decay with the inverse of the radius for large radius and so do the associated solutions.
\item The modified spherical Bessel functions of the first and second kind, $i_l^+$ and $i_l^-$ respectively grow exponentially for large radius and so do the associated solutions.
\end{itemize}
Hence, the only solutions that are well defined and bounded in all of
spacetime arise from the spherical Bessel functions of the first kind
and for $E^2\ge m^2$.

It is sometimes useful to change the basis of the space of
solutions of (\ref{eq:diffbessel}) and use ordinary or modified
spherical Bessel functions of the third kind instead. Concretely, the
spherical Bessel functions of the third kind are defined as,
\be
 h_l=j_l+\im n_l,\quad\text{and}\quad \overline{h}_l=j_l-\im n_l.
\ee
These are complex and have asymptotic behaviours for $z\to\infty$
given by
\be
 h_l(z)\to\im^{-l-1} \frac{e^{\im z}}{z}
  +\mathcal{O}\left(\frac{1}{z^2}\right),\qquad
 \overline{h}_l(z)\to\im^{l+1} \frac{e^{-\im z}}{z}
  +\mathcal{O}\left(\frac{1}{z^2}\right),
\ee
which means that the corresponding solutions (\ref{eq:clkg})
are outgoing ($h_l$) and incoming ($\overline{h}_l$) waves. We also
note that $h_l$ and $\overline{h}_l$ have no zeros on the positive
real axis. The modified spherical Bessel functions of the third kind
are the linear combinations $i_l^+ + \im i_l^-$ and $i_l^+ -\im
i_l^-$. Note that the first of these shows exponential decay at large
radius while the second one shows exponential growth. Of interest to
us will be only the first one, which for $z\to\infty$ behaves as
\be
 i_l^+(z)+\im\, i_l^-(z)\to \im^{-l-2} e^{-z}
 \left(\frac{1}{z}+\mathcal{O}\left(\frac{1}{z^2}\right)\right) .
\ee
Note that this function is either real or imaginary (depending on
$l$).

In order to allow a unified treatment of the cases $E^2> m^2$ and
$E^2< m^2$, we make the following definitions,
\begin{gather}
 a_l(E,r)\defeq
 \begin{cases}
  j_l(r \sqrt{E^2-m^2}) & \text{if}\, E^2>m^2,\\
  i_l^+(r \sqrt{m^2 - E^2}) & \text{if}\, E^2<m^2,
 \end{cases}\qquad\text{and}\qquad
 b_l(E,r)\defeq
 \begin{cases}
  n_l(r \sqrt{E^2-m^2}) & \text{if}\, E^2>m^2,\\
  i_l^-(r \sqrt{m^2 - E^2}) & \text{if}\, E^2<m^2,
 \end{cases}\label{eq:ab}\\
 \text{as well as}\qquad
 c_l(E,r)\defeq a_l(E,r)+\im\, b_l(E,r),
 \qquad\text{and}\qquad
 p\defeq
 \begin{cases}
 \sqrt{E^2-m^2} & \text{if}\, E^2>m^2,\\
 \im\sqrt{m^2-E^2} & \text{if}\, E^2<m^2.
 \end{cases}
\end{gather}

\subsubsection{Quantum theory}

We now turn to the quantum theory. The first object of interest is the
state space associated with a hypercylinder of radius $R$. Since we
are dealing with the Schr\"odinger representation this should be the
space of complex functions on the space of field configurations on
$\R\times S^2_R$. Naively, one would allow essentially any real valued
function as a configuration. However, it was shown in \cite{Oe:KGtl}
that it makes sense to restrict to such configurations which extend to
bounded classical solutions in all of space-time. These configurations
where called \emph{physical configurations}. Recalling the above
discussion of solutions induced by different types of Bessel functions,
this implies a restriction 
of the Fourier expansion of a configuration in the temporal
direction to energy eigenvalues $E$ such that $E^2 > m^2$. In terms of
the particle spectrum it means that we restrict to the
spectrum appearing also in the standard context of equal-time
hypersurfaces where the square $p^2=E^2-m^2$ of the momentum is
non-negative. While we shall keep this restriction for asymptotic
states when considering the S-matrix, we will see that the interacting
theory requires the consideration
of more general configurations and corresponding states in intermediate
calculations. We will thus extend the relevant structures of
\cite{Oe:KGtl} here for that case.

Recall that there are two versions of each hypersurface -- one for
each possible orientation. In the case of the hypercylinder we can
think of the orientation as choosing either the \emph{inside} or the
\emph{outside} of the hypercylinder. To distinguish these, superscripts
$O$ (for outside) and $I$ (for inside) were used in
\cite{Oe:KGtl}. Here we shall only require the outside version of
state spaces and hence do not write any explicit superscript. In
particular, it is clear that the induced orientation of the
hypercylinder as the boundary of the solid hypercylinder is the
outside one. We call this state space $\cH_R$. Hence the amplitude function $\rho_{R}$ on a solid
hypercylinder of radius $R$ is evaluated with an outside state
$\psi\in\cH_R$,
\be
 \rho_{R}(\psi)=\int \xD \varphi\, \psi(\varphi) Z_{R}(\varphi) .
\ee
Here, $Z_R$ is the propagator (\ref{prop}) for $M$ the solid
hypercylinder.

For the free Klein-Gordon theory we call this propagator
$Z_{R,0}$ and its path integral expression can be evaluated
in the same way as for the standard propagator (\ref{eq:proppint}). We
shift the integration by a classical solution $\phi_\text{cl}$
matching the boundary configuration $\varphi$ at radius $R$ to get
\be
 Z_{R,0}(\varphi)=N_{R,0}\, e^{\im S_{R,0}(\phi_\text{cl})}
 ,\quad\text{with}\quad
 N_{R,0}=\int_{\phi|_R=0} \xD\phi\, e^{\im S_{R,0}(\phi)} .
 \label{eq:normr}
\ee
Note that we can rewrite the action (\ref{eq:freeact}) on a classical
solution as a boundary integral. Concretely, on the hypercylinder this
is
\be
 S_{R,0}(\phi_\text{cl})=-\frac{1}{2}\int\xd t\,\xd\Omega\,
  R^2 \phi_\text{cl}(t,R,\Omega) (\partial_r
  \phi_\text{cl})(t,R,\Omega) .
\ee
The classical solution in question must be well defined inside the
solid hypercylinder. This means it can be expanded in the form
(\ref{eq:clkg}) with $f_l=a_l$, i.e., using ordinary and
modified spherical Bessel functions of the first kind, which are regular everywhere.
Since solutions in the interior are essentially in one-to-one
correspondence to boundary
configurations we can relate the two via
\be
\phi_\text{cl}(t,r,\Omega)= \frac{a_l(E,r)}{a_l(E,R)}
\varphi(t,\Omega) .
\label{eq:classbr}
\ee
This allows us to read off the propagator,
\be
 Z_{R,0}(\varphi) = N_{R,0}  \, \exp \left( -\frac{1}{2}
\int \xd t\, \xd \Omega \, \varphi(t,\Omega) \,
\im R^2 \frac{a_l'(E,R)}{a_l(E,R)} \,  \varphi(t,\Omega)\right),
\label{eq:propR}
\ee
where $a_l'$ is the derivative of $a_l$ with respect to $r$ and $R^2
\frac{a_l'(E,R)}{a_l(E,R)}$ is to be understood as an operator via the
mode decomposition of the field. This generalizes the corresponding
result of \cite{Oe:KGtl} to include non-physical
configurations.\footnote{We remark that the inclusion of non-physical
configurations also modifies the normalization factor $N_{R,0}$ as
compared to \cite{Oe:KGtl}, in
spite of the innocent looking expression (\ref{eq:normr}). However,
the explicit form of this and other similar normalization factors is
not of importance here. They can easily be computed using the same
method as in \cite{Oe:KGtl}.}

We will be interested also in a different propagator, namely the one
associated with the region bounded by two hypercylinders of different
radii, say $R$ and $\hat{R}>R$. Again, we can evaluate the path integral
by shifting with a classical solution. However, the class of
solutions to be considered is now larger. In particular, we need not
insist that a solution is well defined at the time axis, since the
latter is not part of the region we consider. Concretely, we
also need to admit ordinary and modified spherical Bessel functions of
the second kind, meaning that in the expansion (\ref{eq:clkg}) we need
not only admit $f_l=a_l$, but also $f_l=b_l$.
Extending the result of \cite{Oe:KGtl} (and using slightly different
conventions for $\delta$ and $\sigma$) the propagator is,
\be
Z_{[R,\hat{R}],0}(\varphi,\hat{\varphi}) = N_{[R,\hat{R}],0}  \,
\exp \left( -\frac{1}{2} \int \xd t\, \xd \Omega \, 
\begin{pmatrix}\varphi & \hat{\varphi} \end{pmatrix} W_{[R,\hat{R}]}  
\begin{pmatrix} \varphi \\ \hat{\varphi} \end{pmatrix}\right),
\label{propRR}
\ee
with
\be
W_{[R,\hat{R}]} = \frac{\im}{\delta_l(E, R, \hat{R})}
\begin{pmatrix} R^2 \sigma_l(E, \hat{R}, R) & - \frac{1}{p}\\
- \frac{1}{p} & \hat{R}^2 \sigma_l(E,R,\hat{R}) \end{pmatrix}.
\ee
The function $\delta_l$ and $\sigma_l$ are to be understood as
operators defined as
\begin{gather}
\delta_l(E, R, \hat{R})
 = a_l(E,R) b_l(E,\hat{R}) - b_l(E,R) a_l(E,\hat{R}) ,\\
\sigma_l(E, R, \hat{R})
 = a_l(E,R) b_l'(E,\hat{R}) - b_l(E,R) a_l'(E,\hat{R}) .
\end{gather}
Again, the derivative refers to the second argument.
Note that both $\delta_l$ and $\sigma_l$ are always
real for $E^2>m^2$ and imaginary for $E^2<m^2$.

We do not repeat here the calculation of \cite{Oe:KGtl} to obtain the
vacuum wave function, but just import the result, extended to
non-physical configurations by analytic continuation,\footnote{Note
  that the vacuum for $E^2<m^2$ is determined here by analytically
continuing the \emph{outside} vacuum state. Instead analytically
continuing the \emph{inside} vacuum state would lead to a different
vacuum. These two choices correspond to the ambiguity described for
$E^2>m^2$ in \cite{Oe:KGtl}.}
\be
\psi_{R,0}(\varphi)= C_R \exp \left( -\frac{1}{2} \int \xd t \,
\xd \Omega \, \varphi(t,\Omega) (B_R  \varphi)(t,\Omega) \right).
\label{eq:vacuumr}
\ee
Here $C_R$ is a normalization factor and $B_R$ denotes the family
of operators indexed by the radius $R$ given by
\be
B_R= \frac{1 - \im
 p R^2 (a_l(E,R)a_l'(E,R)+b_l(E,R)b_l'(E,R))}{p(a_l^2(E,R)+b_l^2(E,R))}
= -\im R^2 \frac{c_l'(E,R)}{c_l(E,R)} .
\ee
Note that for $E^2<m^2$ this operator becomes purely imaginary.

\subsection{Coherent states on the hypercylinder}
\label{sec:cs-hyprc}

We use the following decomposition of field configurations on the
hypercylinder,
\be
 \varphi(t,\Omega)=\int_{-\infty}^\infty \xd E\sum_{l,m}
 \varphi_{l,m}(E) e^{-\im E t}
 Y_l^m(\Omega),\qquad
 \varphi_{l,m}(E)=\frac{1}{2\pi}\int \xd t\,\xd\Omega\,
 \varphi(t,\Omega) e^{\im E t} Y_l^{-m}(\Omega) . 
\ee
We define the family of states
on the hypercylinder of radius $R$,
\be
\psi_{R,\eta}(\varphi) = K_{R,\eta} \,
\exp\left( \int_{|E|\ge m} \xd E \sum_{l,m} \eta_{l,m}(E) \,
\varphi_{l,m}(E) \right) \, \psi_{R,0}(\varphi),
\label{cshyperc}
\ee
parametrized by the complex functions
$\eta_{l,m}(E)$. Here, $K_{R,\eta}$ is a
normalization factor that will be determined later and $\psi_{R,0}$ is the
vacuum state (\ref{eq:vacuumr}). We show in the following that these
states are coherent states in complete analogy to the usual coherent
states on equal-time hypersurfaces: They form a complete set
of states, remain coherent under ``evolution'' etc. (See also the end of
Appendix~\ref{sec:transcoh} for a justification of the definition (\ref{cshyperc})
in terms of creation operators.)
The restriction of $\eta$ to the range $E^2>m^2$ means that we
restrict to a dependence on physical configurations. Sometimes we consider
$\eta_{l,m}$ evaluated on any real value $E$. Then we define
$\eta_{l,m}(E)=0$ if $E^2<m^2$.

The inner product (\ref{eq:iprod}) of coherent states is
\be
\langle \psi_{R,\eta'} | \psi_{R,\eta} \rangle
= \overline{K_{R,\eta'}} K_{R,\eta} \int \xD \varphi \,
\exp\left( \int_{|E|\ge m} \xd E \sum_{l,m} \left(\overline{\eta'_{l,m}(E)}\,
\overline{\varphi_{l,m}(E)}+ \eta_{l,m}(E) \,
\varphi_{l,m}(E) \right)\right) |\psi_{R,0}(\varphi)|^2.
\ee
To evaluate this we notice from (\ref{eq:vacuumr}) that,\footnote{Note
that the vacuum is defined on unphysical configurations as well, but
these do not appear here explicitly because $B_R+\overline{B_R}=0$ if
$E^2<m^2$.}
\be
|\psi_{R,0}(\varphi)|^2 = |C_R|^2 \exp \left( -\int_{|E|\ge m} \xd E \sum_{l,m}
\varphi_{l,m}(E)
\frac{2\pi}{p |h_l(pR)|^2} \varphi_{l,-m}(-E)\right) .
\ee
Since $\varphi$ is real we have
$\overline{\varphi_{l,m}(E)}=\varphi_{l,-m}(-E)$ and,
\begin{multline}
\langle \psi_{R,\eta'} | \psi_{R,\eta} \rangle =
\overline{K_{R,\eta'}} K_{R,\eta} |C_R|^2 \int \xD \varphi \,
\exp\left( \int_{|E|\ge m} \xd E \sum_{l,m} \left(
 \left(\overline{\eta'_{l,-m}(-E)}
  + \eta_{l,m}(E)\right) \varphi_{l,m}(E) \right. \right. \\
- \left. \left. \varphi_{l,m}(E)
\frac{2\pi}{p |h_l(pR)|^2}\varphi_{l,-m}(-E)
\right)\right).
\end{multline}
The integral in $\varphi$ can be evaluated with the usual
technique of shifting the integration variable. In particular, we
implement for $E^2>m^2$ the shift
\be
 \varphi_{l,m}(E) \rightarrow \varphi_{l,m}(E) +
\frac{p |h_l(pR)|^2}{4 \pi} \, \left(
\overline{\eta'_{l,m}(E)} +\eta_{l,-m}(-E) \right) .
\ee
$\varphi_{l,m}(E)$ remains unshifted for $E^2<m^2$.
We obtain the
following expression for the inner product,
\be
\langle \psi_{R,\eta'} | \psi_{R,\eta}\rangle =
\overline{K_{R,\eta'}} \, K_{R,\eta} 
\exp\left( \int_{|E|\ge m} \xd E \sum_{l,m}
  \frac{p |h_l(pR)|^2}{8 \pi}
 \left(  \overline{\eta'_{l,m}(E)} +
 \eta_{l,-m}(-E)\right) \left(\eta_{l,m}(E)+
 \overline{\eta'_{l,-m}(-E)} \right)\right).
\ee

We fix the factors $K_{R,\eta}$ such that the coherent states are
normalized. As is easy to see, this is achieved by setting
\be
K_{R,\eta} = \exp \left(- \int_{|E|\ge m} \xd E \sum_{l,m}
 \frac{p |h_l(pR)|^2}{8 \pi}  \left(\eta_{l,m}(E) \eta_{l,-m}(-E)
+|\eta_{l,m}(E)|^2 \right) \right),
\label{eq:ntilde}
\ee
which yields,
\be
\langle \psi_{R,\eta'} | \psi_{R,\eta} \rangle
= \exp \left(\int_{|E|\ge m} \xd E \sum_{l,m} \frac{p |h_l(pR)|^2}{4
  \pi} \left( \eta_{l,m}(E)
\overline{\eta'_{l,m}(E)} - \frac{1}{2} | \eta_{l,m}(E)|^2 -
\frac{1}{2}|{\eta'}_{l,m}(E)|^2 \right) \right).
\ee

The coherent states satisfy the completeness
relation 
\be
D^{-1} \int \xd \eta \, \xd \overline{\eta} \,
|\psi_{R,\eta}\rangle \langle \psi_{R,\eta}| = I,
\label{com}
\ee
where the constant $D$ is given by
\be
D= \int \xd \eta \, \xd \overline{\eta} \,
\exp \left(-  \int_{|E|\ge m} \xd E \sum_{l,m} \,
 \frac{p |h_l(pR)|^2}{4 \pi} \, |\eta_{l,m}(E)|^2\right).
\label{D}
\ee
We can show the correctness of the completeness relation (\ref{com})
inserting it in an inner product between two coherent states,
\begin{multline}
\langle \psi_{R,\eta}| \psi_{R,\eta'} \rangle = \langle
\psi_{R,\eta}|D^{-1} \int \xd \eta'' \, \xd \overline{\eta''} \,
|\psi_{R,\eta''}\rangle \langle \psi_{R,\eta''}|  \psi_{R,\eta'} \rangle \\
=D^{-1} \exp \left(\int_{|E|\ge m} \xd E \sum_{l,m}
 \frac{p |h_l(pR)|^2}{4 \pi} \left(  - \frac{1}{2} | \eta_{l,m}(E)|^2 -
\frac{1}{2}|{\eta'}_{l,m}(E)|^2 \right) \right) \times \\
   \int \xd \eta'' \, \xd \overline{\eta''} \, \exp \left(\int_{|E|\ge m} \xd E
   \sum_{l,m} \frac{p |h_l(pR)|^2}{4 \pi} \left(  \overline{\eta_{l,m}(E)}
   \eta''_{l,m}(E)+ \overline{\eta''_{l,m}(E)} \eta'_{l,m}(E) -
   |{\eta''}_{l,m}(E)|^2 \right) \right) .
\end{multline}
In order to calculate the integral we shift the integration variables
by the quantities
\be
 \eta''_{l,m}(E) \rightarrow  \eta''_{l,m}(E) +  \eta'_{l,m}(E), \  \
 \text{and} \  \  \overline{\eta''_{l,m}(E)} \rightarrow
 \overline{\eta''_{l,m}(E)} +  \overline{\eta_{l,m}(E)}.
\ee
So we obtain
\begin{multline}
\langle \psi_{R,\eta}| \psi_{R,\eta'} \rangle 
= \exp \left(\int_{|E|\ge m} \xd E \sum_{l,m} \frac{p |h_l(pR)|^2}{4
  \pi} \left(  \eta'_{l,m}(E) \overline{\eta_{l,m}(E)}
- \frac{1}{2} | \eta_{l,m}(E)|^2 - \frac{1}{2}|{\eta'}_{l,m}(E)|^2
\right) \right) \times \\
 D^{-1}  \int \xd \eta'' \, \xd \overline{\eta''} \, \exp
 \left(-\int_{|E|\ge m} \xd E \sum_{l,m} \frac{p |h_l(pR)|^2}{4 \pi}
 \, |{\eta''}_{l,m}(E)|^2  \right) .
\end{multline}
On the right-hand side of the above formula we recognize in the first
line the expression of the inner product $\langle \psi_{R,\eta}|
\psi_{R,\eta'} \rangle$, and using expression (\ref{D}) the second
line is equal to 1. This completes the proof of the completeness relation
(\ref{com}).

We also calculate the amplitude for a coherent state on a solid
hypercylinder. Thus, we have to evaluate the integral
\be
\rho_{R,0}(\psi_{R,\eta}) = \int \xD \varphi \, \psi_{R, \eta}
(\varphi) \, Z_{R,0} (\varphi),
\ee
where the propagator $Z_{R,0}$ is given by (\ref{eq:propR}). This yields,
\be
\rho_{R,0}(\psi_{R,\eta}) = K_{R,\eta}\, C_R \, N_{R,0} \int \xD \varphi
\, \exp\left( \int_{-\infty}^\infty \xd E \sum_{l,m} \left(
    \eta_{l,m}(E) \, \varphi_{l,m}(E) -
    \varphi_{l,m}(E) \, \frac{\pi}{p\, c_l(E,R) a_l(E,R)} \,
    \varphi_{l,-m}(-E) \right) \right).
\label{eq:rho1}
\ee
We shift the integration variable so that the mixed term generated by
the second term in the exponential cancels the term
$\eta_{l,m}(E) \, \varphi_{l,m}(E)$. The shift only concerns physical
configurations and is given by
\be
\varphi_{l,m}(E) \rightarrow \varphi_{l,m}(E) + \frac{p\, h_l(pR)
  j_l(pR)}{2 \pi} \eta_{l,-m}(-E).
\ee
This yields
\be
\rho_{R,0}(\psi_{R,\eta}) = \exp \left( \int_{|E|\ge m} \xd E \sum_{l,m}
 \frac{p}{8 \pi} \left( h_l^2(pR) \eta_{l,m}(E) \eta_{l,-m}(-E)
  - |h_l(pR)|^2 |\eta_{l,m}(E)|^2 \right)
\right) .
\label{eq:rho2}
\ee

The natural analogue of time evolution for the hypercylinder geometry
is radial evolution in space. The evolution of a coherent state from a
hypercylinder of radius $\hat{R}$ to a hypercylinder of radius $R<
\hat{R}$ is determined by the propagator (\ref{propRR}) via
\be
\psi_{R,\eta}(\varphi) = \int \xD \hat{\varphi} \,
\psi_{\hat{R},\hat{\eta}}(\hat{\varphi}) \,
 Z_{[R,\hat{R}],0}(\varphi, \hat{\varphi}).
\ee
(The equation for $R>\hat{R}$ is analogous, but with inside states.)
As is to be expected, a coherent state remains a coherent state under
radial evolution. Indeed, the relation between $\eta$ at $R$ and
$\hat{\eta}$ at $\hat{R}$ turns out to be given by the equation
\be
\eta_{l,m}(E) = \hat{\eta}_{l,m}(E) \, \frac{h_l(p\hat{R})}{h_l(pR)} .
\label{eq:revol}
\ee

The polynomial representation (discussed in
Appendix~\ref{sec:transcoh} mainly for the equal-time hyperplane case)
allows to expand coherent states in terms of multi-particle states,
\begin{multline}
 \psi_{R,\eta}=\exp\left(-\int_{|E|\ge m} \xd E\sum_{l,m}
 \frac{p |h_l(pR)|^2}{8\pi}|\eta_{l,m}(E)|^2\right)\\
 \sum_{n=0}^\infty \frac{1}{n!}
 \int_{|E_1|\ge m} \xd E_1\sum_{l_1,m_1}\cdots
 \int_{|E_n|\ge m} \xd E_n\sum_{l_n,m_n}\,
 \eta_{l_1,m_1}(E_1)\cdots\eta_{l_n,m_n}(E_n)\,
 \psi_{R,(E_1,l_1,m_1),\dots,(E_n,l_n,m_n)}
\label{eq:cohexpand2}
\end{multline}
Here, $\psi_{R,(E_1,l_1,m_1),\dots,(E_n,l_n,m_n)}$ denotes the state
with $n$ particles with the given quantum numbers. Note that our
conventions for multi-particle states are different here from those of
\cite{Oe:KGtl}. For simplicity, we take here the sign of the energy
not to be a separate quantum number and also use a different
normalization. The normalization is fixed by writing down the
one-particle wave function which is,
\be
 \psi_{R,E,l,m}(\varphi)=\varphi_{l,m}(E)\psi_{R,0}.
\label{eq:1pwf}
\ee
We also note the inner product between a coherent state and a
multi-particle state,
\begin{multline}
 \langle \psi_{R,(E_1,l_1,m_1),\dots,(E_n,l_n,m_n)}|
 \psi_{R,\eta} \rangle =\exp\left(-\int_{|E|\ge m} \xd E\sum_{l,m}
 \frac{p |h_l(pR)|^2}{8\pi}|\eta_{l,m}(E)|^2\right)\\
 \eta_{l_1,m_1}(E_1)\cdots\eta_{l_n,m_n}(E_n)\,
 \frac{p_1|h_{l_1}(p_1 R)|^2}{4\pi}\cdots
 \frac{p_n|h_{l_n}(p_n R)|^2}{4\pi} .
\label{eq:cohnip2}
\end{multline}

\section{Asymptotic amplitude on the hypercylinder}
\label{sec:hsmatrix}

Suppose we are interested in a scattering process where interactions
are not negligible at any time, such as a stationary process. A
basic assumption underlying the standard treatment of scattering
processes via the S-matrix as presented in Section~\ref{sec:smatrix}
is then violated. However, suppose that at the same time interactions
can be neglected as we go far away in space from the interaction
center. Recalling Section~\ref{sec:hypercyl} it is obvious that the
hypercylinder geometry is well suited to describe this
situation. Concretely, we consider amplitudes for free states on the
hypercylinder of given radius $R$ with interactions switched on
inside, i.e., for radius $r<R$. We then take the asymptotic limit of
this amplitude for $R\to\infty$.

In this section we show that this
asymptotic amplitude exists and compute it in three different
cases. First we consider the free Klein-Gordon theory, then we add a
source field and finally we consider general interactions via
functional methods.
The derivation in this section proceeds substantially in parallel to
the one of the standard S-matrix in Section~\ref{sec:smatrix}. We use
the coherent states defined in Section~\ref{sec:cs-hyprc}.

\subsection{Free theory}

The first step we need to take in order to make sense of asymptotic amplitudes is
to switch to the interaction picture. Recall that the analogue of time
evolution in the conventional picture is now radial evolution in
space. Hence, the interaction picture means that we identify states on
different hypercylinders if they are related by radial evolution in the free theory.
For coherent states this radial evolution is given by equation
(\ref{eq:revol}). Since the product $\xi_{l,m}(E)\defeq h_l(pR)
\eta_{l,m}(E)$ is preserved under radial evolution a good way to define
interaction picture coherent states on the hypercylinder is by
inserting this relation into (\ref{cshyperc}). We get the wave
functions,
\be
\psi_{R,\xi}(\varphi) = K_{R,\xi} \, \exp\left( \int_{|E|\ge m} \xd E
\sum_{l,m} \frac{\xi_{l,m}(E)}{h_l(pR)} \, \varphi_{l,m}(E) \right) \,
\psi_{R,0}(\varphi),
\label{eq:cshyperc}
\ee
depending on complex functions $\xi_{l,m}(E)$.
The normalization factor can be calculated from
(\ref{eq:ntilde}),
\be
K_{R,\xi} = \exp \left(- \int_{|E|\ge m} \xd E \sum_{l,m} \frac{p}{8 \pi}
  \left(\frac{\overline{h}_l(pR)}{h_l(pR)}
\xi_{l,m}(E) \xi_{l,-m}(-E) + |\xi_{l,m}(E)|^2 \right) \right) .
\label{eq:ipnormh}
\ee

The amplitude of the interaction picture coherent state for the solid
hypercylinder is obtained from (\ref{eq:rho2}),
\be
\rho_{R,0}(\psi_{R,\xi}) = \exp \left( \int_{|E|\ge m} \xd E
\sum_{l,m} \frac{p}{8 \pi} \left(  \xi_{l,m}(E) \xi_{l,-m}(-E)
 - |\xi_{l,m}(E)|^2 \right)
\right).
\label{eq:freeampl}
\ee
By construction this expression is independent of the radius
$R$. Taking the limit $R\to\infty$ is hence trivial and we can write
down the asymptotic amplitude $\cS_0:\cH_\infty\to\C$ of the free
theory,
\be
 \cS_0(\psi_\xi)=\lim_{R\to\infty}\rho_{R,0}(\psi_{R,\xi})
 = \exp \left( \int_{|E|\ge m} \xd E \sum_{l,m} \frac{p}{8
  \pi} \left(  \xi_{l,m}(E) \xi_{l,-m}(-E) - |\xi_{l,m}(E)|^2 \right)
 \right) .
\label{eq:smfree2}
\ee

\subsection{Theory with source}

We now turn to the Klein-Gordon theory with an additional source field
$\mu$, given by the action (\ref{eq:actionsrc}). The spacetime region
of interest is now the solid hypercylinder of radius $R$. We assume
that the source field vanishes outside the solid hypercylinder, i.e.,
for radius $r\ge R$.

The path integral (\ref{prop}) determining the relevant field
propagator $Z_{R,\mu}(\varphi)$ is evaluated as usual by shifting the
integration variable by a solution $\phi_\text{cl}$ that matches the
boundary data. $\phi_\text{cl}$ is defined inside the hypercylinder,
i.e., for radius $r\le R$. It equals $\varphi$ at radius $R$ which we
write as $\phi_\text{cl}|_R=\varphi$. The propagator is then
\be
Z_{R,\mu} (\varphi) =N_{R,\mu}  e^{\im S_{R,\mu}(\phi_\text{cl})}.
\ee
The normalization factor is given by
\be
N_{R,\mu} = \int_{\left. \phi \right|_R=0} \xD\phi \; e^{\im S_{R,\mu}(\phi)}.
\ee
Recall that $\phi_\text{cl}$ can be expanded in the form
(\ref{eq:clkg}) with $f_l=a_l$ and the relation (\ref{eq:classbr})
between solutions and boundary configurations. This allows us to rewrite
the source term of the action (\ref{eq:actionsrc}) evaluated on the
classical solution as
\be
\int_{\R\times B^3_R} \xd^4 x  \, \mu(x) \, \phi_\text{cl}(x)
 = \int_{-\infty}^\infty
 \xd E \sum_{l,m}
\varphi_{l,m}(E) \frac{2\pi}{a_l(E,R)}  M_{l,-m}(-E) ,
\ee
where the quantity $M_{l,m}(E)$ is defined as
\be
M_{l,m}(E) \defeq \int_0^\infty \xd r \; r^2 \; a_l(E,r) \;
\mu_{l,m}(E,r).
\label{eq:defM}
\ee
$\mu_{l,m}(E,r)$ are the modes of the source field $\mu$ in the basis of the
spherical harmonics via the expansion
\be
\mu(t,r,\Omega) = \int_{-\infty}^\infty \xd E\,
\sum_{l,m} \mu_{l,m}(E,r) \,
e^{-\im E t} \, Y_l^{m}(\Omega) .
\label{eq:mudec}
\ee
Then the propagator takes the form
\be
Z_{R,\mu} (\varphi) = \frac{N_{R,\mu}}{N_{R,0}} Z_{R,0} (\varphi) \exp
\left( \int_{-\infty}^\infty \xd E \sum_{l,m} \varphi_{l,m}(E)
 \frac{2\pi \im}{a_l(E,R)}  M_{l,-m}(-E) \right).
\ee
This expression is the analogue of (\ref{eq:propsource}). The
amplitude (\ref{amplitude}) of the theory with source for a coherent
state (\ref{eq:cshyperc}) on the solid hypercylinder is thus,
\be
\rho_{R,\mu}(\psi_{R,\xi})=  \frac{N_{R,\mu}}{N_{R,0}} \int \xD
\varphi \, \psi_{R,\xi}(\varphi) \, \exp\left(
\int_{-\infty}^\infty \xd E
  \sum_{l,m} \varphi_{l,m}(E) \frac{ 2\pi  \im}{a_l(E,R)}
  M_{l,-m}(-E) \right) \, Z_{R,0} (\varphi).
\label{a1}
\ee
We write the coherent state wave function and the
propagator explicitly to get,
\begin{multline}
\rho_{R,\mu}(\psi_{R,\xi}) = K_{R,\xi} C_R N_{R,\mu}
\int \xD\varphi
\exp\left(
\int_{-\infty}^\infty \xd E
\sum_{l,m} \left(\frac{\xi_{l,m}(E)}{h_l(pR)} \, \varphi_{l,m}(E)
 \right. \right.\\
 + \left.\left. \varphi_{l,m}(E) \frac{ 2\pi  \im}{a_l(E,R)}
  M_{l,-m}(-E) - \varphi_{l,m}(E) \;\frac{\pi}{p\,
  c_l(E,R)a_l(E,R)}\;\varphi_{l,-m}(-E) \right)\right) .
\end{multline}
We eliminate the cross term between $\varphi$ and $M$, but only
for unphysical configurations with $E^2<m^2$ via the shift,
\be
\varphi_{l,m}(E) \rightarrow \varphi_{l,m}(E) + \im\, p\, c_l(E,R) M_{l,m}(E).
\ee
Note that this shift is real. That is, the shifted field remains real
(in position space). The corresponding shift for physical
configurations would be complex. We arrive at,
\begin{multline}
\rho_{R,\mu}(\psi_{R,\xi}) =  K_{R,\xi} \frac{N_{R,\mu}}{N_{R,0}} 
\int \xD\varphi
\exp\left(
\int_{|E|\ge m} \xd E
\sum_{l,m} \left(\frac{\xi_{l,m}(E)}{h_l(pR)} \,
\varphi_{l,m}(E)
 + \varphi_{l,m}(E) \frac{ 2\pi  \im}{j_l(pR)} M_{l,-m}(-E)
\right) \right.\\
 - \left.
\int_{-m}^m \xd E
  \sum_{l,m} M_{l,m}(E) \frac{\pi\, p\,
    c_l(E,R)}{a_l(E,R)} M_{l,-m}(-E) \right)
 \psi_{R,0}(\varphi) Z_{R,0}(\varphi).
\end{multline}
In order to deal with the remaining cross term between $\varphi$ and
$M$ for physical configurations, we shift the coherent state. That is,
we define a new coherent state via,
\be
\tilde{\xi}_{l,m}(E) \defeq {\xi}_{l,m}(E) + 2 \pi\im \,
\frac{h_l(pR)}{j_l(pR)} \, M_{l,-m}(-E).
\ee
This yields,
\be
\rho_{R,\mu}(\psi_{R,\xi}) = \frac{K_{R,\xi}}{K_{R,\tilde{\xi}}}\,
\frac{N_{R,\mu}}{N_{R,0}}
 \,\rho_{R,0}(\psi_{R,\tilde{\xi}})
\exp\left(-\int_{-m}^m \xd E
  \sum_{l,m} M_{l,m}(E) \frac{\pi\, p\,
    c_l(E,R)}{a_l(E,R)} M_{l,-m}(-E) \right)
\label{eq:a2}
\ee
Substituting the expression of the free transition amplitude
(\ref{eq:freeampl}) and the normalization factor (\ref{eq:ipnormh})
for the shifted coherent state $\psi_{R,\tilde{\xi}}$, we arrive at
\begin{multline}
\rho_{R,\mu}(\psi_{R,\xi}) = K_{R,\xi}
\frac{N_{R,\mu}}{N_{R,0}}\exp \left( \int_{-\infty}^\infty \xd E
\sum_{l,m} \left( - M_{l,m}(E) \frac{\pi\,p\, c_l(E,R)}{a_l(E,R)}
M_{l,-m}(-E) \right. \right.\\
+ \left. \left.  \im p \, \xi_{l,m}(E) M_{l,m}(E) + \xi_{l,m}(E)
\frac{p\, a_l(E,R)}{4 \pi c_l(E,R)} \xi_{l,-m}(-E) \right) \right).
\end{multline}
Note that the term quadratic in $M$ arises as a combination of the
unphysical part in (\ref{eq:a2}) and a corresponding physical part
coming from the shifted coherent state.

Substituting the expression of the normalization factor $K_{R,\xi}$
given in (\ref{eq:ipnormh}) we obtain
\begin{multline}
\rho_{R,\mu}(\psi_{R,\xi})=  \frac{N_{R,\mu}}{N_{R,0}}\exp \left(
\int_{-\infty}^\infty \xd E \sum_{l,m} \left[\frac{p}{8 \pi} \left(
  \xi_{l,m}(E) \xi_{l,-m}(-E) - |\xi_{l,m}(E)|^2 \right)
  \right. \right.\\
-  \left. \left.   M_{l,m}(E) \frac{\pi\,p\, c_l(E,R)}{a_l(E,R)}
M_{l,-m}(-E) + \im p\, \xi_{l,m}(E) M_{l,m}(E) \right] \right) .
\label{a3}
\end{multline}
We recognize in the first line the amplitude of the coherent state in
the case of the free theory (\ref{eq:freeampl}), so
\be
\rho_{R,\mu}(\psi_{R,\xi}) =  \rho_{R,0}(\psi_{R,\xi})
\frac{N_{R,\mu}}{N_{R,0}} \, \exp \left( \int_{-\infty}^\infty \xd 
E \sum_{l,m} \left[ - M_{l,m}(E) \frac{\pi\,p\, c_l(E,R)}{a_l(E,R)}
  M_{l,-m}(-E) + \im p\, \xi_{l,m}(E) M_{l,m}(E) \right] \right)
\label{eq:rhoS1SS}.
\ee
The second term in the exponential can be rewritten as follows,
\be
\int_{|E|\ge m} \xd E \sum_{l,m}
p\,\xi_{l,m}(E) M_{l,m}(E)
 =\int \xd^4 x \,  \mu(x) \hat{\xi}(x),
\label{eq:ximu}
\ee
with $\hat{\xi}$ given by
\be
\hat{\xi}(t,r,\Omega)\defeq \int_{|E|\ge m} \xd
E\sum_{l,m}\frac{p}{2\pi}\xi_{l,m}(E) j_l(pr) e^{\im E t}
Y^{-m}_l(\Omega) .
\label{eq:classcoh2}
\ee
This implies that we have a one-to-one correspondence between coherent
states parametrized by functions $\xi$ and complex solutions
$\hat{\xi}$ of the Klein-Gordon equation. Note that
the restriction to physical configurations with $E^2<m^2$ precisely
corresponds to restricting solutions to be globally bounded.

We rewrite the first term in the exponential appearing in
(\ref{eq:rhoS1SS}) as,
\be
 \exp\left(-\int_{-\infty}^\infty
 \xd E \sum_{l,m} M_{l,m}(E) \frac{\pi\,p\, c_l(E,R)}{a_l(E,R)}
  M_{l,-m}(-E)\right)
 =\exp\left(\frac{\im}{2}\int \xd^4 x\, \mu(x)\beta(x) \right),
\label{eq:betafac2}
\ee
with $\beta$ a solution of the Klein-Gordon equation given by,
\be
 \beta_{l,m}(E,r)=\im\, p\, a_l(E,r)\frac{c_l(E,R)}{a_l(E,R)} \, M_{l,m}(E)
\label{eq:beta2}
\ee
where we expand,
\be
\beta(t,r,\Omega) = \int_{-\infty}^\infty \xd E \sum_{l,m}
\beta_{l,m}(E,r) e^{-\im Et} \, Y_l^m(\Omega).
\ee

In analogy to the case for equal-time hyperplanes in equation
(\ref{eq:propnormsrc})
the normalization factor
$N_{R,\mu}$ can be related to the normalization factor
$N_{R,0}$. Explicitly,
\be
\int_{\left. \phi \right|_{R}=0} \xD\phi \; e^{\im
  S_{R,\mu}(\phi)} = \exp\left(\frac{\im}{2}\int \xd^4
x\,\mu(x)\alpha(x)\right)
\int_{\left. \phi \right|_{R}=0} \xD\phi \; e^{\im
  S_{R,0}(\phi)},
\ee
which implies,
\be
 \frac{N_{R,\mu}}{N_{R,0}}=\exp\left(\frac{\im}{2}\int \xd^4
x\,\mu(x)\alpha(x)\right),
\label{eq:alphafac2}
\ee
where $\alpha$ satisfies the inhomogeneous Klein-Gordon equation with
vanishing boundary conditions at radius $R$,
\be
(\Box + m^2) \alpha = \mu, \  \ \text{and} \  \ \left. \alpha \right|_R=0.
\ee
It will be convenient to express the exponential factor appearing in
(\ref{eq:alphafac2}) in momentum space,
\be
\exp\left(\frac{\im}{2}\int \xd^4 x\,\mu(x)\alpha(x)\right) =
\exp\left(\im \pi \int_{-\infty}^{\infty} \xd E \sum_{l,m}
 \int_0^\infty \xd r\, r^2
\,\mu_{l,-m}(-E,r) \alpha_{l,m}(E,r) \right),
\ee
where $\alpha_{l,m}(E,r)$ are the modes in the expansion
\be
\alpha(t,r,\Omega) = \int_{-\infty}^\infty \xd E \sum_{l,m}
\alpha_{l,m}(E,r) e^{-\im Et} \, Y_l^m(\Omega).
\label{alpha}
\ee
The inhomogeneous Klein-Gordon equation satisfied by $\alpha$ induces
a differential equation for the modes $\alpha_{l,m}(E,r)$. The
solution results to be equal to
\be
\alpha_{l,m}(E,r) = p\,a_l(E,r) \left(N_{l,m}(E,r) - N_{l,m}(E)+
\frac{b_l(E,R)}{a_l(E,R)} M_{l,m}(E) \right) - p\, b_l(E,r)
M_{l,m}(E,r),
\ee
where in addition to (\ref{eq:defM}) we define,
\begin{gather}
N_{l,m}(E,r) \defeq \int_0^{r} \xd s \, s^2 \, b_l(E,s) \mu_{l,m}(E,s), \label{eq:defNr} \\
M_{l,m}(E,r) \defeq \int_0^{r} \xd s \, s^2 \, a_l(E,s) \mu_{l,m}(E,s), \label{eq:defMr} \\
N_{l,m}(E) \defeq \int_0^{\infty} \xd s \, s^2 \, b_l(E,s) \mu_{l,m}(E,s). \label{eq:defN}
\end{gather}
Notice that for $r>R$, $N_{l,m}(E,r)=N_{l,m}(E)$ and $M_{l,m}(E,r)=M_{l,m}(E)$, because the source components $\mu_{l,m}$ vanish outside the radius $R$.

Combining the factors (\ref{eq:betafac2}) and (\ref{eq:alphafac2})
means that we have to sum $\alpha$ and $\beta$, resulting in
\be
\gamma_{l,m}(E,r) \defeq \alpha_{l,m}(E,r) + \beta_{l,m}(E,r)
 = p\,a_l(E,r) \left(N_{l,m}(E,r) -  N_{l,m}(E) + \im M_{l,m}(E)\right) - p\, b_l(E,r)  M_{l,m}(E,r) .
\label{eq:gamma2}
\ee
Since $\beta$ is a solution of the homogeneous equation, $\gamma$ satisfies the same inhomogeneous equation as $\alpha$, but with different boundary conditions. In particular, for $r>R$, $\gamma$ becomes
\be
\gamma_{l,m}(E,r) \bigg|_{r>R} =  \im\, p \, M_{l,m}(E) \, c_l(E,r).
\label{eq:gammabdy2}
\ee
Now, recall from Section~\ref{sec:hypercyl} that for $E^2>m^2$ the function $c_l$ becomes $h_l$ and generates outgoing waves. For $E^2<m^2$ the function $c_l(E,r)$ becomes $i_l^+ +\im\, i_l^-$ and generates solutions exponentially decaying with the radius. Hence, these are the boundary conditions that $\gamma$ satisfies and that in fact determine it uniquely. This is to be compared with the $\gamma$ obtained in (\ref{eq:gamma}) which is determined by the Feynman boundary conditions (\ref{eq:gammabdy}).

It will be useful to express $\gamma$ in a different way using the definitions (\ref{eq:defM}), (\ref{eq:defNr}), (\ref{eq:defMr}), (\ref{eq:defN}), and (\ref{eq:theta}),
\be
\gamma_{l,m}(E,r) = \im \, p \int_0^{\infty} \xd s  \, s^2 \, \mu_{l,m}(E,s)  \left\{ \theta(r-s) c_l(E,r) \, a_l(E,s) + \theta(s-r) c_l(E,s) \, a_l(E,r) \right\} .
\ee
In position space this is
\be
\gamma(t,r,\Omega) = \im \int_{-\infty}^\infty \xd E\,
\sum_{l,m} \int_0^{\infty} \xd s  \, s^2 \,  p\, \mu_{l,m}(E,s) \, e^{-\im E t} \, Y_l^{m}(\Omega)\,   \left\{ \theta(r-s) c_l(E,r) \, a_l(E,s) + \theta(s-r) c_l(E,s) \, a_l(E,r) \right\}.
\ee
Substituting the inverse of the transformation (\ref{eq:mudec}) we get,
\begin{multline}
\gamma(t,r,\Omega) = \frac{\im}{2\pi} \int_{-\infty}^\infty \xd E\,
\sum_{l,m} \int_0^{\infty} \xd s  \, s^2 \,  p\,  \int \xd t' \xd \Omega' \, e^{\im  Et'} \, \overline{Y_l^{m}(\Omega')} \, \mu(t',s,\Omega') \, e^{-\im E t} \, Y_l^{m}(\Omega) \times \\
\times  \left\{ \theta(r-s) c_l(E,r) \, a_l(E,s) + \theta(s-r) c_l(E,s) \, a_l(E,r) \right\}.
\end{multline}
The sum over $l,m$ can be perform using the formula (\ref{eq:partialwaveGreen}) of Appendix~\ref{sec:appB}.

In the terms of the associated Green function,
\be
\gamma(x) = \int \xd ^4 x' G(x,x') \mu(x'),
\label{eq:green2}
\ee
where
\be
G(t,x,t',x')=\int_{-\infty}^\infty \xd E  \frac{1}{8\pi^2} \frac{e^{\im p |x-x'|+\im   E(t-t')}}{|x-x'|}.
\ee
This Green function can be written in a more familiar form
\be
\int_{-\infty}^\infty \xd E  \frac{1}{8\pi^2} \frac{e^{\im p |\underline{x}-\underline{x}'|+\im E(t-t')}}{|\underline{x}-\underline{x}'|} = -\frac{1}{(2 \pi)^4} \int \frac{e^{- \im  k({x}-{x}')}}{k^2 -m^2 +  \im \epsilon } \xd^4{k}.
\ee
The right-hand side is the standard integral representation of the Feynman propagator of the massive scalar field. Hence $G=G_F$ and the $\gamma$ of this section is identical to the $\gamma$ of Section~\ref{sec:theorysrc}. This implies the surprising fact that for a bounded source the spatially asymptotic boundary conditions (\ref{eq:gammabdy2}) are \emph{equivalent} to the usual temporally asymptotic Feynman boundary conditions (\ref{eq:gammabdy}). Hence we can write the product of $\gamma$ and $\mu$ arising in the combination of (\ref{eq:betafac2}) and (\ref{eq:alphafac2}) as
\be
\frac{\im}{2}\int \xd^4 x\, \mu(x) \gamma(x)= \frac{\im }{2}\int \xd^4 x \, \xd^4 x' \, \mu(x) G_F(x,x')\mu(x').
\label{eq:gammamu}
\ee
We insert (\ref{eq:ximu}), (\ref{eq:betafac2}) and
(\ref{eq:alphafac2}) into (\ref{eq:rhoS1SS}) and use (\ref{eq:gammamu})
to obtain for the amplitude,
\be
\rho_{R,\mu}(\psi_{R,\xi})=  \rho_R(\psi_{R,\xi}) \, \exp\left(\im
\int \xd^4 x \, \hat{\xi}(x) \mu(x)\right)
\exp\left(\frac{\im }{2}\int \xd^4 x \,
\xd^4 x' \, \mu(x) G_F(x,x')\mu(x')\right).
\label{eq:amplrsrc}
\ee
We notice that in this expression
no explicit dependence on the radius $R$ is present, as long as the
support of the source field $\mu$ lies completely within this radius.
So we can take
the limit $R \rightarrow \infty$, lift the restriction on the support
of $\mu$, and interpret the result as the asymptotic amplitude in the
case of a source interaction,
\be
\cS_\mu(\psi_\xi)= \cS_0(\psi_{\xi})\,\exp\left(\im \int \xd^4x\,
\hat{\xi}(x)\mu(x)\right)\exp\left(\frac{\im}{2}\int\xd^4 x\,\xd^4 x'\,
\mu(x)G_F(x,x')\mu(x')\right) .
 \label{eq:smsrc2}
\ee
This is to be compared to (\ref{eq:smsrc}).

\subsection{General interactions}

The transition amplitude in the case of a general interaction
(\ref{eq:actionpot}) can be
computed applying the same functional derivative techniques used in
Section~\ref{sec:generalint}. However, the region where we initially
assume the interaction to vanish is different, i.e.,
\be
V((t, x), \phi(t,x)) = 0, \  \ \hbox{if} \  \ |x|\ge R.
\label{eq:vvanishr}
\ee
Inserting (\ref{eq:actionfunc}) into the path integral of the
propagator (\ref{prop}) where the region $M$ is the solid
hypercylinder of radius $R$, we observe that we can move the exponential
containing the functional derivative out of the integral to the
front. We can
repeat all steps identically as for the theory with source, but with
the functional derivative term in front. Hence, the
amplitude can be read off from (\ref{eq:amplrsrc}) in combination
with (\ref{eq:actionfunc}),
\be
\rho_{R,V}(\psi_{R}) =  \exp\left(\im\int\xd^4 x\,
V\left(x,-\im\frac{\partial}{\partial \mu(x)}\right)\right)
\rho_{R,\mu}(\psi_{R})\bigg|_{\mu=0} .
\label{eq:aint2}
\ee
In this expression a restriction to coherent states is not necessary,
so we have written it for general states.
We now recall that the transition amplitude (\ref{eq:amplrsrc})
does not actually depend on the radius $R$, but is
identical to the asymptotic amplitude of the theory with
source. Hence, expression
(\ref{eq:aint2}) also does not depend on $R$ and the
limit $R\to\infty$ is trivial. We can lift the
restriction (\ref{eq:vvanishr}) on $V$ and obtain the asymptotic
amplitude,
\be
\cS_{V}(\psi) =  \exp\left(\im\int\xd^4 x\,
V\left(x,-\im\frac{\partial}{\partial \mu(x)}\right)\right)
\cS_{\mu}(\psi)\bigg|_{\mu=0} .
\label{eq:smint2}
\ee
This is to be compared to (\ref{eq:smint}).

\section{Equality of asymptotic amplitudes}
\label{sec:compstates}

Although the hypercylinder geometry we used in
Section~\ref{sec:hsmatrix} is rather different from the usual
equal-time hyperplanes, we obtain asymptotic amplitudes that look very
similar to the standard S-matrix expressions.
While the similarity
between the expressions (\ref{eq:smint}) and (\ref{eq:smint2}) is
unsurprising, that of the underlying asymptotic amplitudes with
source, (\ref{eq:smsrc}) and (\ref{eq:smsrc2}), is striking. For
example, the same (Feynman) propagator appears in both
expressions. As we have seen this is due to a non-trivial equivalence
between boundary conditions of the inhomogeneous Klein-Gordon
equation. Namely, the Feynman boundary condition (\ref{eq:gammabdy}) at
past and future temporal infinity is equivalent to an outgoing wave
boundary condition (\ref{eq:gammabdy2}) at spatial infinity.

From a physical point of view the similarity should not
surprise us. If we want to describe a process that is bounded
both in space and time (i.e., such that interactions may be neglected
at large spatial and temporal distance) the hypercylinder geometry
should serve as well as the standard one. Indeed, we show in this
section that the asymptotic amplitudes in the two settings are
\emph{identical}, if the correct identification of temporal and
spatial asymptotic states is performed.

In the standard setting, let us call the asymptotic state space in the
infinite past $\cH_1$. Correspondingly, we call the asymptotic state
space in the infinite
future $\cH_2$. The tensor product $\cH_1\tens\cH_2^*$ is the total
Hilbert space of states at spatial infinity. (The dualization of the
future state space occurs because initial states are ket-states while
final ones are bra-states.) The S-matrix (\ref{eq:smint}) is then a
linear map from the Hilbert space $\cH_1\tens\cH_2^*$ to the complex
numbers. In the hypercylinder setting, we call the asymptotic state
space at infinite radius $\cHc$.

Identifying asymptotic states at
temporal and at spatial infinity now translates to an isomorphism of
Hilbert spaces $\cH_1\tens\cH_2^*\to\cHc$.
We can read off this isomorphism by comparing 
(\ref{eq:smsrc}) and (\ref{eq:smsrc2}) as follows.
Recall that we have a one-to-one correspondence
between coherent states $|\psi_{\eta_1}\rangle\tens\langle\psi_{\eta_2}|
\in\cH_1\tens\cH_2^*$ and complex classical
solutions $\hat{\eta}$ in spacetime via
(\ref{eq:classcoh}). Similarly, we have a
one-to-one correspondence between coherent states $\psi_{\xi}\in\cHc$ and
complex classical solutions $\hat{\xi}$ in spacetime via
(\ref{eq:classcoh2}). For the spatially asymptotic amplitude
(\ref{eq:smsrc2}) to agree with the S-matrix (\ref{eq:smsrc}) we
obviously need to identify the classical solutions, i.e.,
$\hat{\xi}=\hat{\eta}$. Otherwise the $\mu$-dependence would be
different in the two expressions. It remains to check that this
identification also makes the free amplitudes (\ref{eq:smfree}) and
(\ref{eq:smfree2}) equal.

Consider a bounded complex solution $\hat{\xi}$ of the Klein-Gordon
equation in Minkowski spacetime. This solution corresponds via its
decomposition into positive and negative
frequency components (\ref{eq:classcoh}) which we shall write also as
$\hat{\xi}=\hat{\xi}_+ + \hat{\xi}_-$
to a pair of coherent states whose free S-matrix (\ref{eq:smfree}),
i.e.\ inner product (\ref{eq:sproduct-cs}) is
\begin{multline}
 \langle \psi_{\overline{\xi_-}}|\cS_0|\psi_{\xi_+}\rangle=
 \exp\left(\int \frac{\xd^3 k}{(2\pi)^3 2 E}\left(\xi_+(k) \xi_-(k)
 -\frac{1}{2}|\xi_+(k)|^2-\frac{1}{2}|\xi_-(k)|^2 \right)\right)\\
 =\exp\left(\int\xd^3 x\, \left(2\hat{\xi}_+(t,x)(\omega\hat{\xi}_-)(t,x)
  -\hat{\xi}_+(t,x) (\omega\overline{\hat{\xi}_+})(t,x)
  -\hat{\xi}_-(t,x) (\omega\overline{\hat{\xi}_-})(t,x)
  \right)\right)
\end{multline}
In the last line the time $t$ is arbitrary.
Inserting the decomposition of $\hat{\xi}_\pm$ in terms of spherical
harmonics and Bessel functions,
\be
\hat{\xi}_{\pm}(t,r,\Omega) = \int_m^{\infty} \xd E \sum_{l,m}
\frac{p}{2 \pi}  \,
j_l(pr)  \, Y^{-m}_l (\Omega) \, \xi_{l,m}(\mp E) \, e^{\mp \im Et},
\label{eq:xipm}
\ee
yields after integration over $x$,
\be
\langle \psi_{\overline{\xi_-}} |\cS_0| \psi_{\xi_+} \rangle
= \int_{|E|\ge m}  \xd E \sum_{l,m} \frac{p}{8 \pi} \left(
\, \xi_{l,m}(E) \xi_{l,-m}(- E)  - |\xi_{l,m}(E)|^2 \right)
=\cS_0(\psi_\xi),
\ee
the free S-matrix on the hypercylinder (\ref{eq:smfree2}), as required.

We turn now to look at the isomorphism $\cH_1\tens\cH_2^*\to\cHc$ in
terms of multi-particle states. To this end we use the completeness
relation for coherent states together with the formulas from
Sections~\ref{sec:coherent-state} and \ref{sec:cs-hyprc}
relating coherent states to multi-particle states. However, we have to
transform these formulas first into the interaction picture. For the
standard setting of equal-time hyperplanes this transformation is
trivial and relations (\ref{eq:cohcompl}) as well as
(\ref{eq:cohexpand}) and (\ref{eq:cohnip}) simply remain the same in
the interaction picture.
In the hypercylinder setting the situation is different. Indeed, we
first have to define particle states in the interaction picture. A
suitable definition which we shall use in this section is,
\be \psi_{R,E,l,m}(\varphi)=\frac{\varphi_{l,m}(E)}{h_l(pR)}\psi_{R,0}(\varphi),
\ee
for a one-particle wave function. This is to be contrasted with
(\ref{eq:1pwf}). The definition of multi-particle
wave functions is then fixed in the usual way and the inner product between two $n$-particle states is 
\be
\langle \psi_{(l_1,m_1,E_1),\dots,(l_n,m_n,E_n)} | \psi_{(l_1',m_1',E_1'),\dots,(l_n',m_n',E_n')} \rangle = 
\frac{p_1}{4 \pi} \dots \frac{p_n}{4 \pi}  \delta_{l_1,l_1'} \dots \delta_{l_n,l_n'} \, \delta_{m_1,m_1'} \dots \delta_{m_n,m_n'} \, \delta(E_1 - E_1') \dots \delta(E_n - E_n').
\ee

The completeness relation (\ref{com}) remains true, although with
a modified factor $D$. Indeed, formula (\ref{D}) as well as
(\ref{eq:cohexpand2}) and (\ref{eq:cohnip2}) change in
so far as all factors of $|h_l(pR)|^2$ disappear. We will only need
the completeness relation,
\be
\tilde{D}^{-1} \int \xd \xi \, \xd \overline{\xi} \,
|\psi_{\xi}\rangle \langle \psi_{\xi}| = I,
\label{eq:complip}
\ee
with
\be
\tilde{D}= \int \xd \xi \, \xd \overline{\xi} \,
\exp \left(-  \int_{|E|\ge m} \xd E \sum_{l,m} \,
 \frac{p}{4 \pi} \, |\xi_{l,m}(E)|^2\right),
\label{eq:tildeD}
\ee
and the inner product,
\be
 \langle \psi_{(E_1,l_1,m_1),\dots,(E_n,l_n,m_n)}|
 \psi_{\xi} \rangle =\exp\left(-\int_{|E|\ge m} \xd E\sum_{l,m}
 \frac{p}{8\pi}|\xi_{l,m}(E)|^2\right)\,
 \xi_{l_1,m_1}(E_1)\cdots\xi_{l_n,m_n}(E_n)\,
 \frac{p_1}{4\pi}\cdots
 \frac{p_n}{4\pi} .
\label{eq:ipip}
\ee

To denote a state in $\cH_1\tens\cH_2^*$ with $q$ incoming particles with momenta $k_1,\dots,k_q$ and $n-q$ outgoing particles with momenta $k_{q+1},\dots,k_n$ we write $\psi_{k_1,\dots,k_q | k_{q+1},\dots,k_n}$. In the standard bra-ket notation this translates into,
\be
 \psi_{k_1,\dots,k_q | k_{q+1},\dots,k_n} = |\psi_{k_1,\dots,k_q}\rangle\tens\langle\psi_{k_{q+1},\dots,k_n}|.
\ee
Let now $\hat{\xi}$ be a bounded classical solution in spacetime and $\hat{\xi}_+,\hat{\xi}_-$ the positive/negative frequency components as defined above. The associated coherent state in $\psi_\xi\in\cH_1\tens\cH_2^*$ takes in the standard bra-ket notation the form,
\be
 \psi_\xi=|\psi_{\xi_+}\rangle \tens \langle\psi_{\overline{\xi_-}}|.
\ee
The inner product of a coherent state $\psi_\xi$ with an $n$-particle state $\psi_{k_1,\dots,k_q | k_{q+1},\dots,k_n}$ thus takes the form
\be
 \langle \psi_\xi | \psi_{k_1,\dots,k_q | k_{q+1},\dots,k_n}\rangle
 = \langle \psi_{\xi_+}|\psi_{k_1,\dots,k_q}\rangle
   \langle\psi_{k_{q+1},\dots,k_n}|
   \psi_{\overline{\xi_-}}\rangle
 = C_{\xi_+} C_{\overline{\xi_-}}\, \overline{\xi_+(k_1)}\cdots
 \overline{\xi_+(k_q)}\,\overline{\xi_-(k_{q+1})}\cdots
 \overline{\xi_-(k_n)},
\label{eq:ipdouble}
\ee
where we have used (\ref{eq:cohnip}). It will be useful to reexpress this in terms of the spherical harmonic modes of $\hat{\xi}$. To this end we notice that (\ref{eq:classcoh}) implies,
\be
\xi_\pm(k) =\, \int \xd ^3x \, \left( E \, \hat{\xi}(x,t) \pm \im \dot{\hat{\xi}}(x,t) \right) e^{\pm\im (E t -kx)},  \\
\ee
where a dot indicates the derivative with respect to time. Inserting (\ref{eq:xipm}) and integrating over $x$ yields,
\be
\xi_\pm(k) = 2 \pi  \sum_{l,m} \xi_{l,m}(\mp E) \, \im^{\mp l} \, Y_l^{-m}(\Omega_k),
\ee
where the angle coordinates $\Omega_k$ are given by the direction of the 3-vector $k$. Using this latter relation we rewrite (\ref{eq:ipdouble}) as,
\begin{multline}
\langle \psi_{\xi} |\psi_{k_1,\dots,k_q | k_{q+1},\dots,k_n} \rangle
 =  (-1)^{l_{q+1}+ \cdots + l_n}\left( 2 \pi \right)^n i^{l_1+
 \dots + l_n} \sum_{l_1,m_1} \cdots \sum_{l_n,m_n}
 \overline{\xi_{l_1,m_1}(-E_{k_1})} \cdots
 \overline{\xi_{l_q,m_q}(-E_{k_q})} \times \\
 \overline{\xi_{l_{q+1},m_{q+1}}(E_{k_{q+1}})} \cdots
 \overline{\xi_{l_{n},m_{n}}(E_{k_{n}})} \, 
 Y_{l_1}^{m_1}(\Omega_1) \cdots Y_{l_n}^{m_n}(\Omega_n) \,  \exp
 \left( - \int_{|E|\ge m} \xd E \sum_{l,m} \frac{p}{8 \pi}
 |\xi_{l,m}(E)|^2\right).
\label{eq:innerprod2}
\end{multline}

We now combine the inner products with the completeness relation (\ref{eq:complip}) to an inner product between an $n$-particle state in the standard asymptotic state space $\cH_1\tens\cH_2^*$ and an $n$-particle state in the hypercylinder asymptotic state space $\cHc$,
\be
\langle \psi_{(l_1,m_1,E_1),\dots,(l_n,m_n,E_n)} | \psi_{k_1,\dots,k_q | k_{q+1},\dots,k_n}
\rangle = \tilde{D}^{-1} \int \xd \xi \xd \overline{\xi} \, \langle
\psi_{(l_1,m_1,E_1),\dots,(l_n,m_n,E_n)} |\psi_{\xi} \rangle \langle
\psi_{\xi} |\psi_{k_1,\dots,k_q | k_{q+1},\dots,k_n} \rangle.
\ee
We evaluate this using the expressions (\ref{eq:tildeD}), (\ref{eq:ipip}) and (\ref{eq:innerprod2}) together with the usual method of shifting integration variables to arrive at,
\begin{multline}
\langle \psi_{(l_1,m_1,E_1),\dots,(l_n,m_n,E_n)} | \psi_{k_1,\dots,k_q | k_{q+1},\dots,k_n}
\rangle = (-1)^{l_{q+1}+ \cdots + l_n} \left( 2\pi \right)^n
i^{l_1+ \dots + l_n} \, Y_{l_1}^{m_1} (\Omega_{k_1}) \cdots
Y_{l_n}^{m_n} (\Omega_{k_n}) \times \\ 
\delta(E_1+E_{k_1}) \cdots  \delta(E_q+E_{k_q}) \,
\delta(E_{q+1}-E_{k_{q+1}}) \cdots  \delta(E_n-E_{k_n}).
\end{multline}
We can now write an $n$-particle state in $\cHc$ as a linear combination of $n$-particle states in $\cH_1\tens\cH_2^*$
\begin{multline}
\psi_{(l_1,m_1,E_1),\dots,(l_n,m_n,E_n)} = \frac{(-1)^{l_{q+1}+ \cdots + l_n} 
i^{-l_1- \dots - l_n}}{(8\pi^2)^n} \int \xd \Omega_{p_1} \cdots \int \xd \Omega_{p_n} \, Y_{l_1}^{-m_1} (\Omega_{p_1}) \cdots Y_{l_n}^{-m_n} (\Omega_{p_n}) \, p_1 \cdots p_n \times \\
 \psi_{p_1,\dots,p_q |  p_{q+1},\dots,p_n} .
\label{eq:isom1}
\end{multline}
Reciprocally an $n$-particle state in $\cH_1\tens\cH_2^*$ is a linear combination of $n$-particle states in $\cHc$
\begin{multline}
\psi_{k_1,\dots,k_q | k_{q+1},\dots,k_n} = \frac{ \left( 8\pi^2 \right)^n }{ k_1  \cdots k_n }  \sum_{l_1,m_1} \dots  \sum_{l_n,m_n} (-1)^{l_{q+1}+ \cdots + l_n} i^{l_1+ \dots + l_n}   Y_{l_1}^{m_1} (\Omega_{k_1}) \cdots Y_{l_n}^{m_n} (\Omega_{k_n}) \times \\
\psi_{(l_1,m_{1},-E_{k_1}),\dots,(l_{q},m_{q},-E_{k_q}),(l_{q+1},m_{q+1},E_{k_{q+1}}),\dots,(l_n,m_n,E_{k_n})} .
\label{eq:isom2}
\end{multline}

The decomposition of (the finite radius version of) $\cHc$ as a tensor product of in- and out-state spaces $\cH^-\tens\cH^+$ was already introduced in \cite{Oe:KGtl}. Also, the fact that (with the present conventions) negative energy particles are in-particles while positive energy particles are out-particles was already discussed in that paper as well as in \cite{Oe:timelike}. This already explains partially the formulas (\ref{eq:isom1}) and (\ref{eq:isom2}), especially with respect to the energy quantum numbers. From that perspective, what is new here is that we have constructed the isomorphisms $\cH_1\to\cH^-$ between in-state spaces and $\cH_2^*\to\cH^+$ between out-state spaces. In terms of quantum numbers, this means we can now explicitly convert between the angular momentum quantum numbers $l,m$ in $\cHc$ and the three-momentum directions $\Omega_k$ in $\cH_1$ and $\cH_2$.

\section{Conclusions and Outlook}
\label{sec:conclude}

The present result has immediate significance as a contribution to the GBF program in general, and to the extensibility conjecture specifically. It shows that and how perturbative interacting quantum field theory fits into the GBF for an asymptotic geometry that implements key non-standard features of the GBF. In particular, the geometry in question involves a region (the solid hypercylinder) whose boundary is timelike and connected, both features in contrast to what can be made sense of in the standard formalism. It is worth emphasizing at this point that the asymptotic geometry chosen is less special than it might seem from looking at its finite cousin. In contrast to the finite radius hypercylinder, its infinite radius limit is Poincar\'e invariant and may thus be seen as the limit of all possible timelike hypercylinders. This is analogous, of course, to what happens with the standard geometry. The asymptotic pair of equal-time hyperplanes at infinite initial and final time is Poincar\'e invariant and arises as the limit of all possible pairs of spacelike hyperplanes. The precise meaning of these statements is that the asymptotic amplitudes (\ref{eq:smint}) and (\ref{eq:smint2}) are Poincar\'e invariant. This in turn comes from the fact that the propagator appearing in the underlying expressions (\ref{eq:smsrc}) and (\ref{eq:smsrc2}) is Poincar\'e invariant. This reflects the Poincar\'e invariance of the relevant boundary conditions of the inhomogeneous Klein-Gordon equation. While this invariance is well known for the Feynman boundary conditions (\ref{eq:gammabdy}) it came as a surprise for the spatially asymptotic boundary conditions (\ref{eq:gammabdy2}). Indeed, it follows from the even more surprising fact that both boundary conditions are equivalent (for a bounded source).

Concerning possible geometries for constructing asymptotic amplitudes, one might envision further ones than are different from both the standard one and the hypercylinder geometry. A fairly straightforward situation should be that obtained by parallel \emph{timelike} hyperplanes \cite{Oe:timelike,Oe:KGtl} and the limit of moving these to opposite spatial infinities. Based on our present finding we may say with confidence that the result would be an asymptotic amplitude equivalent to both the standard S-matrix as well as the asymptotic hypercylinder amplitude calculated here. A more interesting case would be to start with a finite region, e.g., a four-ball with its boundary three-sphere, letting the radius go to infinity. This would involve the novel feature of hypersurfaces that have both spacelike and timelike parts. A similarly interesting geometry could be that of a diamond. This would, apart from null hypersurfaces, involve \emph{corners}. (See \cite{Oe:2dqym} for a discussion of corners in the GBF.) Of course, one should expect the asymptotic amplitudes constructed with these types of geometries to be equivalent as well if the extensibility conjecture is valid in some generality.

Apart from the relevance to the GBF program our result has further independent implications. One is related to the fact that there is only one spatial asymptotic state space in the hypercylinder geometry in contrast to the two (initial and final) temporal asymptotic state spaces for the standard S-matrix. Whether a particle in the spatial asymptotic state space is in-coming or out-going (and hence appears under the isomorphism in the initial or final temporal asymptotic state space) is determined by its (energy) quantum numbers. Hence, the usual notion of crossing symmetry, a derived property in conventional treatments of the S-matrix, becomes a tautology in the hypercylinder geometry. To put it the other way round: If crossing symmetry did not hold, the GBF would be in trouble. Indeed, the anticipation of crossing symmetry as a manifest feature of the GBF served as an initial motivation for its development \cite{Oe:catandclock}, independent from quantum gravity considerations.

Let us explore further the new spatially asymptotic amplitude introduced in this paper. We have seen in Section~\ref{sec:compstates} that it is equivalent to the usual S-matrix (when both make sense). What is more, as shown in \cite{Oe:GBQFT,Oe:KGtl} the GBF gives it a full fledged physical interpretation, fit for its description of scattering processes, and independent of the mentioned equivalence. Hence, the spatial asymptotic amplitude may be called ``S-matrix'' with the same justification as the usual one and we will do so from here onwards. However, this does not mean that both S-matrices are equally applicable in all physical situations. Indeed, for the usual S-matrix to make sense, interactions must be negligible at very early and at very late times. In contrast, the spatially asymptotic S-matrix requires that interactions are negligible at large distance from a center. However, the interactions may remain important at all times. One may argue that the latter restriction is more naturally met than the former in many physical situations of interest, such as (almost) stationary processes. One may thus expect a useful role for the new S-matrix for the description of certain processes where the usual S-matrix approach fails.

Generalizing our approach to curved spacetime, the difference between the new S-matrix and the usual one would become even more important. What we mean here with this generalization is the following. One would choose in the spacetime in question a region that shares the characteristic features of the solid hypercylinder (such as its topology and the fact that its boundary is timelike). One would then establish the interaction picture for such hypercylinders under radial ``evolution''. Finally, one would compute the asymptotic amplitude when the radius is taken to infinity. Of course, this would work only in a certain class of spacetimes. For example, a condition would be that space be non-compact. The point is that the class of spacetimes where this would work is different from the one where a conventional S-matrix description works.
For example, the present approach should be applicable to Anti-de~Sitter space, where a conventional S-matrix does not exist because a useful notion of temporal asymptotic state space is lacking. Nevertheless, an ``S-matrix analogue''  (which plays an important role in the conjectured AdS/CFT correspondence \cite{Mal:adscft}) has been constructed for Anti-de~Sitter in a more heuristic way \cite{Gid:smatrixadscft}. The present approach should likely lead at least to a better conceptual underpinning of this construction.
Another interesting example could be a stationary black hole spacetime. Placing a hypercylinder carrying the states outside the horizon naturally avoids having to say anything about the black hole interior within this state space. Unsurprisingly, ideas in this direction are already implicit in approaches to black hole physics. For example, 't~Hooft factorizes a hypothetical black hole S-matrix into three pieces \cite{tHo:smatrixbh}, one of which could be interpreted as corresponding to a hypercylinder amplitude just outside the horizon.

The method of derivation used here for the new as well as for the standard S-matrix merits an additional comment. In contrast to most treatments of the S-matrix we first construct finite interacting (transition) amplitudes explicitly and then take the respective asymptotic limit. This is facilitated by the use of coherent states. For the case of the hypercylinder geometry these were not known previously, so their introduction in Section~\ref{sec:cs-hyprc} constitutes a side result of this paper.

\begin{acknowledgments}

This work was supported in part by CONACyT grants 47857 and 49093.

\end{acknowledgments}

\appendix

\section{Transformation of coherent states into the Schr\"odinger
 representation and polynomial representation}
\label{sec:transcoh}

In order to determine the wave functions of coherent states in the
Schr\"odinger representation it is useful to transform to a
representation which we shall call the \emph{polynomial
representation}.
This representation is related to the holomorphic one, although we
will not discuss this relation here.

We first consider the context of equal-time hyperplanes. Recall that
the wave function of a multi-particle state in the Schr\"odinger
representation is a product of a polynomial with the vacuum wave
function. Say, we have $n$ particles with momenta
$p_1,\dots,p_n$. The wave function of the associated normalized state
takes the form
\be
 \psi_{p_1,\dots,p_n}(\varphi)=P_{p_1,\dots,p_n}(\varphi)\,
 \psi_0(\varphi) .
\ee
$P_{p_1,\dots,p_n}$ is a polynomial of order $n$ in the field
configuration $\varphi$. It is not a monomial, however, but rather a
linear combination of monomials of orders $n$, $n-2$, $n-4$ etc. The
order $n$ component is the product $\check{\varphi}(p_1)\cdots
\check{\varphi}(p_n)$ (in the conventions of \cite{Oe:KGtl}) where
$\check{\varphi}(p)$ is the functional given by Fourier transform,
\be
 \check{\varphi}(p)\defeq 2 E \int\xd^3 x\, e^{\im p x}
 \varphi(x) .
\ee

As the Schr\"odinger representation, the representation we are
going to define consists of a certain space of functions on
instantaneous field configurations. We can thus think of its elements
also as ``wave functions''. To distinguish them from the wave functions
in the Schr\"odinger representation we add a tilde to the
former. Concretely, the transformation $\psi\mapsto\widetilde{\psi}$
between the representations is given by the formula,
\be
 \widetilde{\psi}(\varphi) \defeq |C|^2
 \int \xD \varphi' \frac{\psi(\varphi+\varphi')}{\psi_0(\varphi+\varphi')}
 \exp\left(-\int\xd^3 x\, \varphi'(x) (\omega \varphi')(x) \right),
\label{eq:transpoly}
\ee
where $C$ is the usual normalization constant appearing in the vacuum wave
function (\ref{eq:vacuum}). Note that the inverse transform is given by,
\be
 \psi(\varphi) = |\widetilde{C}|^2\, \psi_0(\varphi)
 \int \xD \varphi'\, \widetilde{\psi}(\varphi+\varphi')
 \exp\left(\int\xd^3 x\, \varphi'(x) (\omega \varphi')(x) \right) ,
\label{eq:invptrans}
\ee
where the normalization factor $\widetilde{C}$ is formally defined via
\be
 |\widetilde{C}|^{-2}\defeq \int \xD\varphi\,
 \exp\left(\int\xd^3 x\, \varphi(x) (\omega \varphi)(x) \right) .
\ee

The defined representation, to which we shall refer as the \emph{polynomial
representation}, has the nice property that the wave function of an
$n$-particle state is simply a monomial of degree $n$,
\be
 \widetilde{\psi}_{p_1,\dots,p_n}(\varphi)
 =\check{\phi}(p_1)\cdots \check{\phi}(p_n) .
\ee
The vacuum wave function is just the unit constant,
$\widetilde{\psi}_0(\varphi)=1$. A creation operator acts by
multiplication, while an annihilation operator acts by derivation,
\be
 (a^\dagger(p)\, \widetilde{\psi})(\varphi)
 =\check{\varphi}(p)\, \widetilde{\psi}(\varphi),\qquad
 (a(p)\, \widetilde{\psi})(\varphi)
 = \int\xd^3 x\, e^{-\im p x}\frac{\delta}{\delta\varphi(x)}\,
 \widetilde{\psi}(\varphi) .
\ee

The simple action of the creation operator makes it easy to work out
the wave function of the coherent state (\ref{eq:coherentstate}) of
Section~\ref{sec:coherent-state} in
the polynomial representation. We obviously get,
\be
 \widetilde{\psi}_\eta(\varphi) = C_\eta \exp \left(
 \int \frac{\xd ^3k}{(2 \pi)^3 2 E} \, \eta(k) \check{\varphi}(k)\right) .
\ee
It remains to apply the inverse transform (\ref{eq:invptrans}) to
obtain the Schr\"odinger representation wave function. We can read off
the field dependent part immediately: Inserting a sum of two fields
into the exponential simply gives a product of exponentials. So, the
field dependence remains exactly the same in the Schr\"odinger
representation (up to the vacuum wave function factor),
\be
 \psi_\eta(\varphi) = K_\eta \exp \left(
 \int \frac{\xd ^3k}{(2 \pi)^3 2 E} \, \eta(k)
 \check{\varphi}(k)\right)
 \psi_0(\varphi) .
\label{eq:cohschr}
\ee
Only the
$\eta$-dependent normalization factor changes. It can be computed with
the usual method of shifting integration variables and the result is
given by (\ref{eq:cohnorm}).

The Schr\"odinger representation on the hypercylinder has a structure
analogous to the one on equal-time hyperplanes \cite{Oe:KGtl}. Indeed,
formulas are structurally identical. This is also true for the polynomial
representation and its relation to the Schr\"odinger
representation. We mention here only the transformation formula analogous
to (\ref{eq:transpoly}) which now reads,
\be
  \widetilde{\psi}(\varphi) \defeq |C_R|^2
 \int \xD \varphi' \frac{\psi(\varphi+\varphi')}{\psi_{R,0}(\varphi+\varphi')}
 \exp\left(-\int_{|E|\ge m} \xd E\sum_{l,m}
 \varphi'_{l,m}(E) \frac{2\pi}{p |h_l(p R)|^2} \varphi'_{l,-m}(-E) \right),
\ee
where $C_R$ is the usual normalization constant appearing in the vacuum
wave function (\ref{eq:vacuumr}).
Hence, if we define a coherent state on the
hypercylinder via creation operators acting on the vacuum in analogy to
(\ref{eq:coherentstate}) the result will necessarily be of a form
analogous to (\ref{eq:cohschr}). This may serve as a justification for
the definition (\ref{eq:cshyperc}) in Section~\ref{sec:cs-hyprc}.

\section{Spherical Harmonics and Spherical Bessel functions}
\label{sec:appB}

The solutions of the eigenvalue problem corresponding to the angular part of the Klein-Gordon equation in spherical coordinates, namely equation (\ref{eq:diffY}), are given by the spherical harmonics denoted $Y^m_l$. These functions are defined through the associated Legendre function $P_l^m$ \cite{AbSt:handbook} via
\be
Y_l^m(\theta, \phi) = \sqrt{\frac{(2l+1)(l-m)!}{4 \pi (l+m)!}} \, P_l^m( \cos \theta) e^{\im m \phi}.
\ee
and satisfy the orthogonality relation
\be
\int \xd \Omega\, Y_l^m \overline{Y_{l'}^{m'}} = \delta_{l,l'} \delta_{m,m'},
\label{ORY}
\ee
where $\int\xd\Omega=4\pi$.\footnote{We differ here slightly from the conventions in \cite{Oe:KGtl}, where $\int\xd\Omega=1$.} Note also, $\overline{Y_{l}^{m}}={Y_{l}^{-m}}$.

The radial part of the Klein-Gordon equation in spherical coordinates, namely equation (\ref{eq:diffbessel}), is solved by the so called \emph{spherical Bessel functions} of the first kind $j_l$ and of the second kind $n_l$ and the \emph{modified spherical Bessel functions} of the first kind $i_l^+$ and of the second kind $i_l^-$. 

The spherical Bessel functions $j_l$ and $n_l$ can be expressed in terms of the ordinary Bessel functions of the first and second kind, $J_l$ and $N_l$ respectively, as
\be
j_l(z) = \sqrt{\frac{\pi}{2z}} J_{l+1/2}(z),\quad\text{and}\quad n_l(z) = \sqrt{\frac{\pi}{2z}} N_{l+1/2}(z).
\ee
Here we use the conventions of \cite{AbSt:handbook} for the ordinary spherical Bessel functions, \emph{but} different conventions for the modified spherical Bessel functions. Indeed we define the modified spherical Bessel functions as the analytic continuation of the ordinary ones,
\be
 i_l^+(z)\defeq j_l(e^{\im \pi/2} z),\quad\text{and}\quad
 i_l^-(z)\defeq n_l(e^{\im \pi/2} z).
 \label{B4}
\ee
Our convention differs by powers of $\im$ from the definitions 10.2.2 and 10.2.3 of \cite{AbSt:handbook}. However, this turns out to be more convenient for our treatment.
We note that $i_l^+$ and $i_l^-$ are always either real or imaginary (depending on $l$) and the product $i_l^+ i_l^-$ is always imaginary.

The Bessel functions satisfy the orthogonality relation
\be
\int_0^{\infty}\xd z \, z \, J_l(\alpha z) J_l(\beta z) = \frac{1}{\alpha } \delta(\alpha - \beta), \quad\text{for}\quad l>-1, \, \alpha>0, \, \beta>0.
\label{eq:ortB}
\ee
An analogous relation is valid for $N_l$. (\ref{eq:ortB}) follows from formulas 6.633.2 of \cite{GR:tables} and 9.7.1 of \cite{AbSt:handbook}. This relation implies for the spherical Bessel functions,
\be
\int_0^{\infty}\xd z \, z^2 \, j_l(\alpha z) j_l(\beta z) = \frac{\pi}{2\alpha^2 } \delta(\alpha - \beta), \quad\text{for}\quad l>-\frac{3}{2}, \, \alpha>0, \, \beta>0.
\label{eq:ortBs}
\ee
The same relation holds for $n_l$.

The Wronskians for the spherical Bessel functions and the modified spherical Bessel functions are
\be
j_l(z) \, n_l'(z) - j_l'(z) \, n_l(z) = \frac{1}{z^2},\quad\text{and}\quad i_l^+(z) \, (i_l^-)'(z) - (i_l^+)'(z) \, i_l^-(z) = - \frac{ \im}{z^2}.
\label{Wronsk}
\ee
where the first equation above corresponds to 10.1.6 of \cite{AbSt:handbook} and the second follows from (\ref{B4}) and 10.2.7 of \cite{AbSt:handbook}. Following the generalized notation for the ordinary and the modified spherical Bessel functions introduced at the end of Section \ref{sec:hypclassical}, formulas (\ref{eq:ab}), from (\ref{Wronsk}) we derive the relation
\be
a_l(E,z) \, b_l'(E,z) - a_l'(E,z) \, b_l(E,z) = \frac{1}{p z^2},  \qquad\text{with}\qquad
 p\defeq
 \begin{cases}
 \sqrt{E^2-m^2} & \text{if}\, E^2>m^2,\\
 \im\sqrt{m^2-E^2} & \text{if}\, E^2<m^2.
 \end{cases}
\ee

The partial wave decomposition of a plane wave is expressed in terms of spherical harmonics and spherical Bessel functions as (see formula (B.105) of \cite{Mes:qm1})
\be
e^{\im k z}= 4 \pi \sum_{l,m} \im^l \, j_l(E_k |z|) \, \overline{Y_l^m(\Omega_k)} \, Y_l^m(\Omega_z),
\ee
where $\Omega_k$ and $\Omega_z$ represent the $\theta, \phi$ directions of the 3-vectors $k$ and $z$ respectively, and $E_k=\sqrt{|k|^2+m^2}$. Using the notation (\ref{eq:ab}) for the spherical Bessel functions, from the above decomposition of a plane wave follows the relation
\be
\frac{e^{\im p \, |z-z'|}}{4 \pi  |z-z'|}  = \im p \sum_{l,m} \, \overline{Y_l^m(\Omega_z)} \,  {Y_l^{m}(\Omega_{z'})} \, \left\{ \theta(|z|-|z'|) c_l(E, |z|) \, a_l(E, |z|) + \theta(|z'|-|z|) c_l(E, |z'|) \, a_l(E, |z|) \right\},
\label{eq:partialwaveGreen}
\ee
see formulas (B.98) and (B.100) of \cite{Mes:qm1}. Notice that $p$ can be real or imaginary.

\bibliography{stdrefs}
\bibliographystyle{amsordx}

\end{document}